\documentclass[lettersize,journal]{IEEEtran}
\UseRawInputEncoding
\usepackage{amsmath}
\DeclareMathOperator{\tr}{tr}
\DeclareMathOperator{\rank}{rank}
\DeclareMathOperator{\diag}{diag}
\usepackage{amssymb}
\usepackage[table,xcdraw]{xcolor}
\usepackage{bm}
\usepackage{color}
\usepackage{graphicx, floatrow}
\usepackage{stfloats}  
\usepackage{float}  
\floatstyle{plaintop}
\restylefloat{table}
\usepackage{subfigure}  
\usepackage{empheq}
\usepackage[linesnumbered,ruled,lined]{algorithm2e}
\usepackage[outdir=./]{epstopdf}
\usepackage{lipsum}
\usepackage{cite}
\usepackage{multirow}
\usepackage{hyperref}
\hypersetup{hidelinks,
	colorlinks=true,
	allcolors=black,
	pdfstartview=Fit,
	breaklinks=true}
\usepackage{fancyhdr}
\usepackage{graphicx}
\SetKwInput{Input}{input}
\SetKwInput{Output}{output}

\usepackage{array}
\newcolumntype{L}[1]{>{\raggedright\let\newline\\\arraybackslash\hspace{0pt}}m{#1}}
\newcolumntype{C}[1]{>{\centering\let\newline\\\arraybackslash\hspace{0pt}}m{#1}}
\newcolumntype{R}[1]{>{\raggedleft\let\newline\\\arraybackslash\hspace{0pt}}m{#1}}
\usepackage{booktabs}
\newcolumntype{I}{!{\vrule width 1.05pt}}

\newtheorem{remark}{Remark}
\IEEEoverridecommandlockouts
\usepackage[acronym]{glossaries-extra}
\setabbreviationstyle[acronym]{long-short}
\newacronym{awgn}{AWGN}{Additive White Gaussian Noise}
\newacronym{ber}{BER}{Bit Error Rate}
\newacronym{bs}{BSs}{Base Stations}
\newacronym{csir}{CSIR}{Channel State Information at Receiver}
\newacronym{csit}{CSIT}{Channel State Information at Transmitter}
\newacronym{csi}{CSI}{Channel State Information}
\newacronym{drsma}{D-RSMA}{Distributed-RSMA}
\newacronym{embb}{eMBB}{Enhanced Mobile Broadband}
\newacronym{fdma}{FDMA}{Frequency Division Multiple Access}
\newacronym{geo}{GEO}{Geostationary Orbit}
\newacronym{gus}{GUs}{GEO UTs}
\newacronym{gu}{GU}{GEO UT}
\newacronym{gw}{GW}{gateway}
\newacronym{iid}{i.i.d}{independent and identically distributed}
\newacronym{leo}{LEO}{Low Earth Orbit}
\newacronym{lhs}{LHS}{Left Hand Side}
\newacronym{lmi}{LMI}{Linear Matrix Inequality}
\newacronym{los}{LoS}{Line-of-Sight}
\newacronym{lus}{LUs}{LEO UTs}
\newacronym{lu}{LU}{LEO UT}
\newacronym{mmf}{MMF}{Max-Min Fairness}
\newacronym{mmtc}{mMTC}{massive Machine Type Communications}
\newacronym{mrt}{MRT}{Maximum Ratio Transmission}
\newacronym{ngeo}{NGEO}{Non-Geostationary}
\newacronym{noma}{NOMA}{Non-Orthogonal Multiple Access}
\newacronym{nr}{NR}{New Radio}
\newacronym{ofdm}{OFDM}{Orthogonal Frequency Division Multiplexing}
\newacronym{oma}{OMA}{Orthogonal Multiple Access}
\newacronym{rhs}{RHS}{Right-Hand Side}
\newacronym{rsma}{RSMA}{Rate Splitting Multiple Access}
\newacronym{sca}{SCA}{Successive Convex Approximation}
\newacronym{sdma}{SDMA}{Space Division Multiple Access}
\newacronym{sdp}{SDP}{Semi-Definite Programming}
\newacronym{sic}{SIC}{Successive Interference Cancellation}
\newacronym{sinr}{SINR}{Signal to Interference plus Noise Ratio}
\newacronym{snr}{SNR}{Signal-to-Noise Ratio}
\newacronym{soc}{SOC}{Second Order Cone}
\newacronym{stin}{STIN}{Satellite-Terrestrial Integrated Network}
\newacronym{svd}{SVD}{Singular Value Decomposition}
\newacronym{uav}{UAV}{Unmanned Aerial Vehicle}
\newacronym{urllc}{URLLC}{Ultra-Reliable and Low-Latency Communications}
\newacronym{uts}{UTs}{User Terminals}
\newacronym{wmmse}{WMMSE}{Weighted Minimum-Mean Square Error}

\begin{document}
\title{Distributed Rate-Splitting Multiple Access for Multilayer Satellite Communications}

\author{
\IEEEauthorblockN{Yunnuo~Xu, Longfei Yin, \IEEEmembership{Student Member, IEEE}, Yijie~Mao, \IEEEmembership{Member, IEEE}, \\Wonjae~Shin, \IEEEmembership{Senior Member, IEEE} and~Bruno~Clerckx, \IEEEmembership{Fellow, IEEE}
 }
\thanks{
\par Y. Xu and L. Yin are with the Communications and Signal Processing Group,
Department of Electrical and Electronic Engineering, Imperial College
London, London SW7 2AZ, U.K. (email: {yunnuo.xu19, longfei.yin17}@imperial.ac.uk).
\par Y. Mao is with the School of Information Science and Technology, ShanghaiTech University, Shanghai 201210, China (email: maoyj@shanghaitech.edu.cn).
\par W. Shin is with the School of Electrical Engineering, Korea University, Seoul 02841, South Korea (email: wjshin@korea.ac.kr).
\par B. Clerckx is with the Department of Electrical and Electronic Engineering at Imperial College London, London SW7 2AZ, UK (email: b.clerckx@imperial.ac.uk).
}
}

\maketitle

\begin{abstract}
	Future wireless networks, in particular, 5G and beyond, are anticipated to deploy dense Low Earth Orbit (LEO) satellites to provide global coverage and broadband connectivity. However, the limited frequency band and the coexistence of multiple constellations bring new challenges for interference management. In this paper, we propose a robust multilayer interference management scheme for spectrum sharing in heterogeneous satellite networks with statistical Channel State Information (CSI) at the Transmitter (CSIT) and Receivers (CSIR).
	In the proposed scheme, Rate-Splitting Multiple Access (RSMA), as a general and powerful framework for interference management and multiple access strategies, is implemented distributedly at Geostationary Orbit (GEO) and LEO satellites, coined Distributed-RSMA (D-RSMA). By doing so, D-RSMA aims to mitigate the interference and boost the user fairness of the overall multilayer satellite system.
	Specifically, we study the problem of jointly optimizing the GEO/LEO precoders and message splits to maximize the minimum rate among User Terminals (UTs) subject to a transmit power constraint at all satellites. A robust algorithm is proposed to solve the original non-convex optimization problem. Numerical results demonstrate the effectiveness and robustness towards network load and CSI uncertainty of our proposed D-RSMA scheme. Benefiting from the interference management capability, D-RSMA provides significant max-min fairness performance gains compared to several benchmark schemes.

\end{abstract}
\vspace{-1mm}
\begin{IEEEkeywords}
	RSMA, max-min fairness, beamforming design, statistical CSIT and CSIR, satellite communication.
	\vspace{-3mm}
\end{IEEEkeywords}

\section{Introduction}
\par \gls{leo} satellite is envisioned as an appealing technique to support and enhance 5G \gls{nr} and beyond-5G communications \cite{22822,38821}. 
It has benefits in enabling network scalability, reinforcing the communication service availability and reliability, and boosting the performance of limited terrestrial networks in un-served/underserved areas \cite{38811,9815078}. 
\gls{leo} constellations formed by \gls{leo} satellites are required to provide worldwide and continuous coverage. Due to the limited frequency resource, however, different satellite constellations should work in the same frequency band, which induces mutual interference and further destabilizes the reliability of space-to-ground links, especially in \gls{geo}-\gls{leo} coexisting scenarios \cite{8928079,9512414}. 

\par To manage the mutual interference, some existing works \cite{9512414,8352660,8928079} study the power controlling-based methods. On the one hand, the authors in \cite{8352660} studied the \gls{geo} and \gls{leo} coexistence system where the \gls{leo} satellites serve as primary and the \gls{geo} satellite acts as secondary. Beam hopping and adaptive power control techniques are implemented at the \gls{geo} satellite to maximize the system throughput and minimize the interference from \gls{geo} to the \gls{leo} network.
On the other hand, the authors of \cite{8928079} and \cite{9512414} considered a case where the \gls{geo} satellite is treated as primary and the \gls{leo} as secondary. \cite{8928079} focused on a single \gls{geo}-single \gls{leo} co-existing system, and the authors formulated a beam power control problem solved by a fractional programming algorithm aiming to maximize the \gls{leo} transmission rate while satisfying the service quality requirement of the \gls{geo} satellite. Further, \cite{9512414} extended the scenario to a single \gls{geo}-multiple \gls{leo} satellites network, a flexible spectrum sharing and cooperative communication method is proposed to mitigate the inter-system interference, where \gls{leo} users are served by multiple \gls{leo} satellites cooperatively.
\par In contrast to the works \cite{9512414,8352660,8928079} that concentrate on power control in non-terrestrial networks, beamforming techniques can also be utilized to coordinate the interference between different satellite systems and improve system performance.
Specifically, \gls{rsma}, relying on linear precoding rate-splitting at the transmitter and \gls{sic} at the receivers, has emerged as a promising multiple access technique for modern multi-antenna networks \cite{10273395,9831440,10038476,9451194}.
\gls{rsma} divides user messages into common and private parts at the transmitter for interference management.
The common parts of the split messages are combined and encoded into a common stream, then decoded by multiple users, while the private parts are encoded separately and decoded by their corresponding users (and treated as noise by co-scheduled users) \cite{7513415,Mao2018,9831440,9461768}. 
There are various structures of \gls{rsma}, such as 1-layer \gls{rsma}, hierarchical \gls{rsma}, and generalized \gls{rsma} \cite{Mao2018}. 
\gls{rsma} has also been shown to be a unified and generalized multiple access schemes that subsumes \gls{sdma}, \gls{noma}, and physical layer multicasting as special cases \cite{8907421,Mao2018}. 
{The gains of \gls{rsma} have been demonstrated in various multi-antenna terrestrial networks in terms of energy efficiency \cite{7738598,8491100}, user fairness \cite{9123680,9991090}, robustness and latency \cite{9970313,9562192,liu2022network,9831048} for a wide range of network loads and channel state conditions.}
\par More recently, building upon the promising gains offered by \gls{rsma}
	, several works extended the application of \gls{rsma} to \gls{uav} communications \cite{9258414}, satellite communications \cite{8491094,9257433,9684855,10045781} and the \gls{stin} \cite{9844445,10266774}.
	Specifically, in \cite{9258414}, \gls{uav}s served users in a unicast manner, and the weighted sum rate was maximized by jointly optimizing \gls{uav} placement and precoders.
\cite{8491094} implemented \gls{rsma} in a two-beam \gls{geo} satellite communication system to mitigate the inter-beam interference and improve the achievable rate region. \cite{9257433} considered an \gls{rsma}-based multibeam multicast network with different \gls{csit}
qualities, in which a \gls{wmmse}-based algorithm was proposed to maximize the \gls{mmf} ergodic rate.
A general \gls{stin} framework was presented in \cite{9844445,10266774}, in which the mutual interference between satellite users and cellular ones is taken into account. The authors designed two \gls{rsma}-based \gls{stin} schemes to suppress interference aiming to maximize the minimum fairness rate among all satellite users and cellular users. Simulation results showed that the proposed scheme achieves higher \gls{mmf} rate than the conventional schemes without \gls{rsma}, illustrating the effectiveness of \gls{rsma} to manage the interference among satellites.
\begin{figure*}[t!]
	\centering
	\includegraphics[scale=0.45]{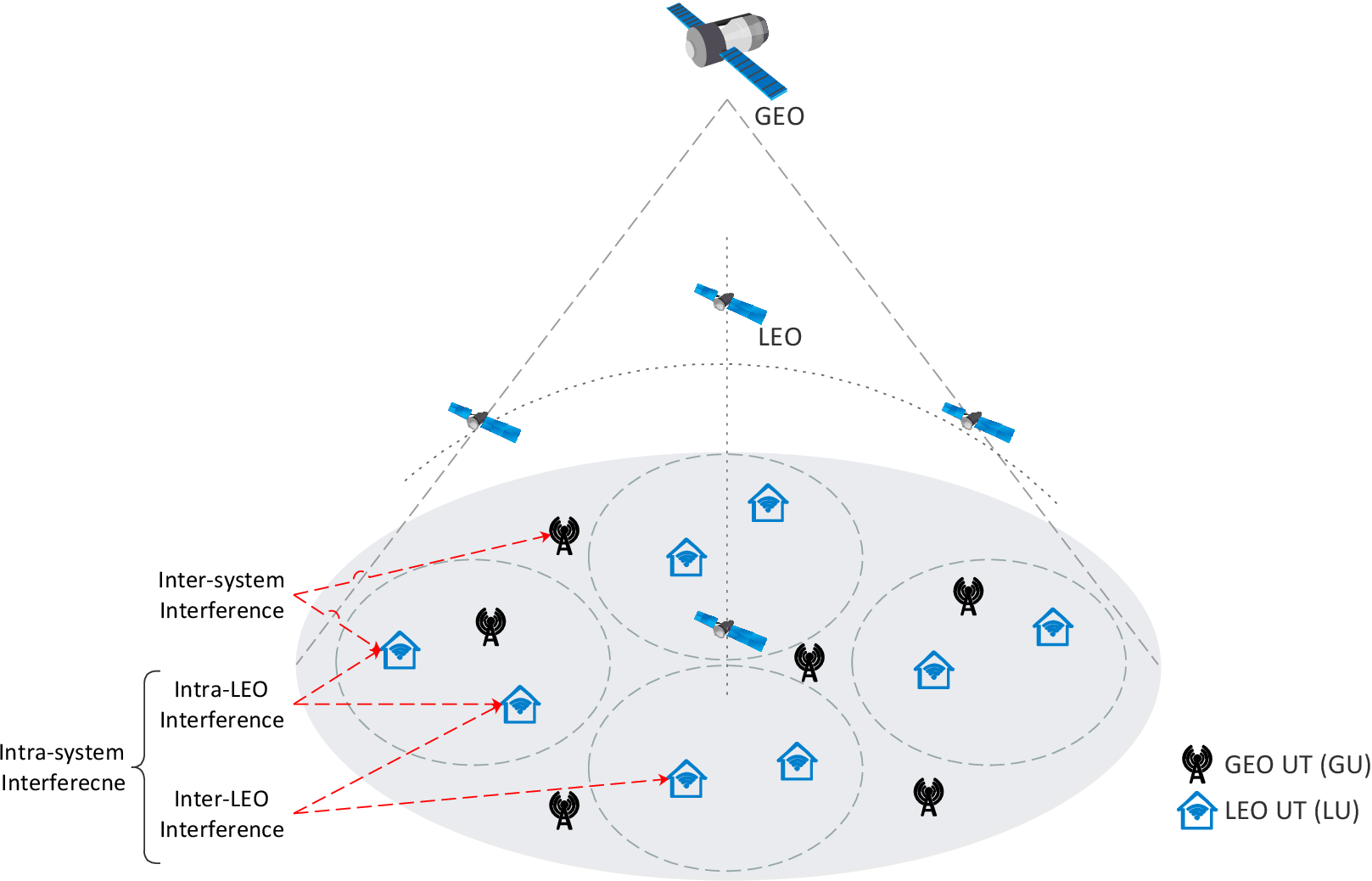}
	\caption{Multilayered satellite system model with $1$ \gls{geo} satellite and $4$ \gls{leo} satellites, $N=6$, $K=8$.}
	\label{fig:sys_illu}
\end{figure*}
\par However, \cite{9258414} focused on \gls{uav} communications with limited coverage. Besides, \gls{uav}s or terrestrial \gls{bs} typically serve users in a unicast manner \cite{9258414, 7875081}, while \gls{geo} typically serves users in a multicast manner with larger coverage. In the considered \gls{geo}-\gls{leo} coexistence networks, we consider the combination of unicast and multicast modes, and the co-channel users served by \gls{leo} nodes suffer from the interference from both \gls{geo} and unintended \gls{leo} satellites.
\cite{8491094,9257433,9844445,10266774} consider \gls{geo}-only or \gls{stin} network and do not investigate multiple orbit constellations co-existence network.
There is limited work which investigated a single-\gls{geo}-single \gls{leo} satellite network \cite{10097680}, where the \gls{geo} satellite adopted \gls{oma} and the \gls{leo} employed \gls{rsma}-based multigroup multicast transmission.
In \cite{10097680}, a \gls{geo} satellite serves as primary network while \gls{leo} operates as secondary network. \gls{rsma} is used to mitigate the interference and improve the sum rate of \gls{leo} users. Nevertheless, \cite{10097680} does not make full use of \gls{rsma}, since it only optimizes the power allocation without beamforming. Besides, it relies on perfect \gls{csit} and \gls{csir}, which lacks practicality in real-world communication situations.

\par Motivated by
$1)$ the severe interference in multilayer satellite networks,
$2)$ the appealing interference management capability of \gls{rsma}, and $3)$ the lack of more comprehensive beamforming and power control strategies in \gls{geo}-\gls{leo} coexistence networks,
we further investigate a multilayer satellite network consisting of a \gls{geo} satellite and multiple \gls{leo} satellites (as shown in Fig. \ref{fig:sys_illu}). \gls{rsma} is implemented across satellites to manage the interference in and between both sub-networks under imperfect \gls{csit} and \gls{csir}. The contributions of this article are summarized as follows:
\begin{enumerate}
	\item We first present a general framework to manage the interference in \gls{geo}-\gls{leo} coexistence networks,  where \gls{rsma} is distributedly implemented, namely \gls{drsma}, at both \gls{geo} and \gls{leo} satellite sub-networks. In this framework, we assume that both satellite systems share the same radio spectrum due to the spectrum scarcity. The gateway is deployed to gather information, allocate resources, and manage interference.
	      The structure of the proposed scheme is designed to encompass the conventional RSMA. This framework differs from prior \gls{rsma} \gls{stin} papers due to the involvement of multiple \gls{leo} satellites, and a distributed implementation strategy of \gls{rsma}, i.e., D-RSMA, at both \gls{geo} and \gls{leo} networks.
	\item It is difficult to acquire the instantaneous \gls{csi} in real-world, however, statistical \gls{csi} varies slower than instantaneous \gls{csi}, which allows it to be relatively accurately and easily captured by both the satellites and \gls{uts}. Therefore, we investigate a robust beamforming design for \gls{drsma} with statistical \gls{csit} and \gls{csir}. To the best of our knowledge, this is the first work on joint \gls{geo}-\gls{leo} beamforming design with imperfect \gls{csit} and \gls{csir}.
	\item Based on the proposed multilayer \gls{drsma} framework, we formulate an \gls{mmf} problem to jointly optimize the beamforming and power allocation of \gls{geo} and \gls{leo} satellites as well as the message splits of \gls{rsma}. Instead of focusing on the performance of one of the sub-networks, and maintaining the quality of service for the other sub-network, the objective of this paper is to maximize the minimum rate among all \gls{uts}.
	\item Due to the nonconvex and mathematical intractable essence of the formulated optimization problem, we transform the original problem into a tractable form by using \gls{sdp} and first-order Taylor approximation, and then a penalty function-based iterative algorithm is proposed to tackle the optimization problem. Numerical results demonstrate that the proposed \gls{drsma} scheme is robust to user deployment and channel uncertainty. \gls{drsma} achieves a \gls{mmf} rate gain over benchmark schemes.
\end{enumerate}
\par The rest of this paper is organized as follows. The system model and multilayer \gls{drsma} scheme are introduced in Section \ref{sec:System_model}.
The \gls{mmf} optimization problem and proposed robust \gls{sdp}-based iterative optimization algorithm are specified in Section \ref{sec:Prob_formu}. 
Simulation results illustrating the effectiveness of our proposed scheme are discussed in Section \ref{sec:Simu_results}, followed by the conclusions in Section \ref{sec:Conclu}. The major variables adopted in the paper are listed in Table \ref{tab:vari_list} for ease of reference.

\par In the remaining sections of this work, matrices, column vectors, and scalars are denoted by boldface uppercase, boldface lowercase, and standard letters, respectively. The operator $(\cdot)^T$ denotes transpose and $(\cdot)^H$ denotes conjugate-transpose. $\circ$ is Hadamard product. $\lambda_{\max}(\cdot)$ denotes the maximum eigenvalue of the matrix. $\mathcal{CN}(\zeta,\sigma^2)$ represents a complex Gaussian distribution with mean $\zeta$ and variance $\sigma^2$. $\text{tr}(\cdot)$ is the trace. $|\cdot|$ is the absolute value and $\|\cdot\|$ is the Euclidean norm. $\mathbb{C}$ denotes the complex space. $|\mathcal{A}|$ is the cardinality of the set $\mathcal{A}$.

\vspace{-1mm}
\section{System model}\label{sec:System_model}
\par We consider a multilayer satellite system with a single \gls{geo} satellite and $M$ \gls{leo} satellites serving $N$ single-antenna \gls{gus} and $K$ single-antenna \gls{lus}, respectively.
\gls{gus} are the \gls{uts} served by the \gls{geo}, and \gls{lus} are the \gls{uts} served by the \gls{leo} satellites.
Fig. \ref{fig:sys_illu} shows a toy example of the system model with $1$ \gls{geo} and $4$ \gls{leo} satellites serving $N=6$ \gls{gus}, and $K=8$ \gls{lus}. 
We denote the set of \gls{leo} satellites as $\mathcal{M}=\{1,\ldots,M\}$. $\mathcal{N}=\{1,\ldots,N\}$ and $\mathcal{K}=\{1,\ldots,K\}$ are the sets of \gls{uts} served by \gls{geo} and \gls{leo}, respectively. \gls{lus} are divided into $M$ groups, and $\mathcal{K}_m$ denotes the set of \gls{uts} under the coverage of $m$-th \gls{leo} satellite. $\bigcup_{m\in\mathcal{M}}\mathcal{K}_i=\mathcal{K}$,
$\mathcal{K}_i\cap\mathcal{K}_j=\emptyset$, $i,\ j\in\mathcal{M},\ i\neq j$. The size of the $m$-th group is $K_m=|\mathcal{K}_m|$ 
, $\forall k\in\mathcal{K}_m$. \gls{geo} and \gls{leo} satellites are equipped with $N_{tg}$ and $N_{tl}$ feeds array fed reflector antennas, respectively. We assume that the \gls{geo} and \gls{leo} satellites operate in the same radio spectrum. The satellites are managed by the \gls{gw}s, which act as a control hub to gather and handle various types of data, apply centralized processing, and oversee the overall system through resource allocation and interference management.
We assume that statistical \gls{csit} and \gls{csir} are available in the network.

\begin{table}[t]
	\caption{Variable List}\vspace{-2mm}
	\label{tab:vari_list}
	\begin{tabular}{IlIlI}
		\bottomrule[1.05pt]
		\hline
		Notation               & \multicolumn{1}{cI}{Definition}               \\ \toprule[1.05pt]
		\bottomrule[1.05pt]
		$M$                    & number of \gls{leo} satellites                \\ \hline
		$N$                    & number of \gls{gus}                           \\ \hline
		$K$                    & number of \gls{lus}                           \\ \hline
		$N_{tg}$               & number of antennas for \gls{geo}              \\ \hline
		$N_{tl}$               & number of antennas for \gls{leo}              \\ \hline
		$\mathbf{h}_{g,n}$     & channels between \gls{geo} and GU-$n$         \\ \hline
		$\mathbf{h}_{g2l,k_m}$ & channels between \gls{geo} and LU-$k_m$       \\ \hline
		$\mathbf{h}_{l,mk_m}$  & channels between \gls{leo}-$m$ and LU-$k_m$   \\ \hline
		$\mathbf{h}_{l2g,mn}$  & channels between \gls{leo}-$m$ and GU-$n$     \\ \hline
		$s_{g,c}$              & common stream from \gls{geo}                  \\ \hline
		$s_{g,d}$              & GUs-designated stream from \gls{geo}          \\ \hline
		$s_{l,m}^{sub}$        & sub-common stream from \gls{leo}-$m$          \\ \hline
		$s_{p,k_m}$            & private stream from \gls{leo}-$m$             \\ \hline
		$\mathbf{w}_c$         & precoders for \gls{geo} common                \\ \hline
		$\mathbf{w}_d$         & precoders for \gls{geo} GUs-designated stream \\ \hline
		$\mathbf{p}_{c,m}$     & precoders for \gls{leo}-$m$ sub-common stream \\ \hline
		$\mathbf{p}_{p,k_m}$   & precoders for \gls{leo}-$m$ private stream    \\ \toprule[1.05pt]
	\end{tabular}
	\vspace{-4mm}
\end{table}

\par Due to the spectrum sharing and co-existence \gls{geo}/\gls{leo} satellites, each user experiences hierarchical multi-user interference. \gls{gus} receive inter-system interference from \gls{leo} satellites, besides, \gls{lus} suffer from inter-\gls{leo} interference from unintended \gls{leo} satellites as well as from inter-system interference from \gls{geo}.
The \gls{geo} and \gls{leo} satellites can exploit various multiple access techniques to manage the interference, such as \gls{rsma}, \gls{sdma}, multicast \cite{7765141} etc.
Inspired by the state-of-the-art \gls{rsma} frameworks and the interference coordination capability of \gls{rsma} \cite{9831440}, we propose a scheme where \gls{rsma} is distributedly implemented at different satellites, which is denoted as \gls{drsma}.
Different from prior literature which typically deploys traditional \gls{rsma} in one network, in this work, the proposed \gls{drsma} implements \gls{rsma} across the networks (i.e., \gls{geo} and \gls{leo} sub-networks), enhancing the ability of interference management.
In this section, we first elaborate the received signal by \gls{gus} and \gls{lus} followed by channel models and imperfect channel model. Then we illustrate the signal processing progress at \gls{gus} and \gls{lus}.

\vspace{-2mm}
\subsection{Transmit and Received Signal}
\begin{figure*}[!t]
	\centering
	\includegraphics[scale=0.67]{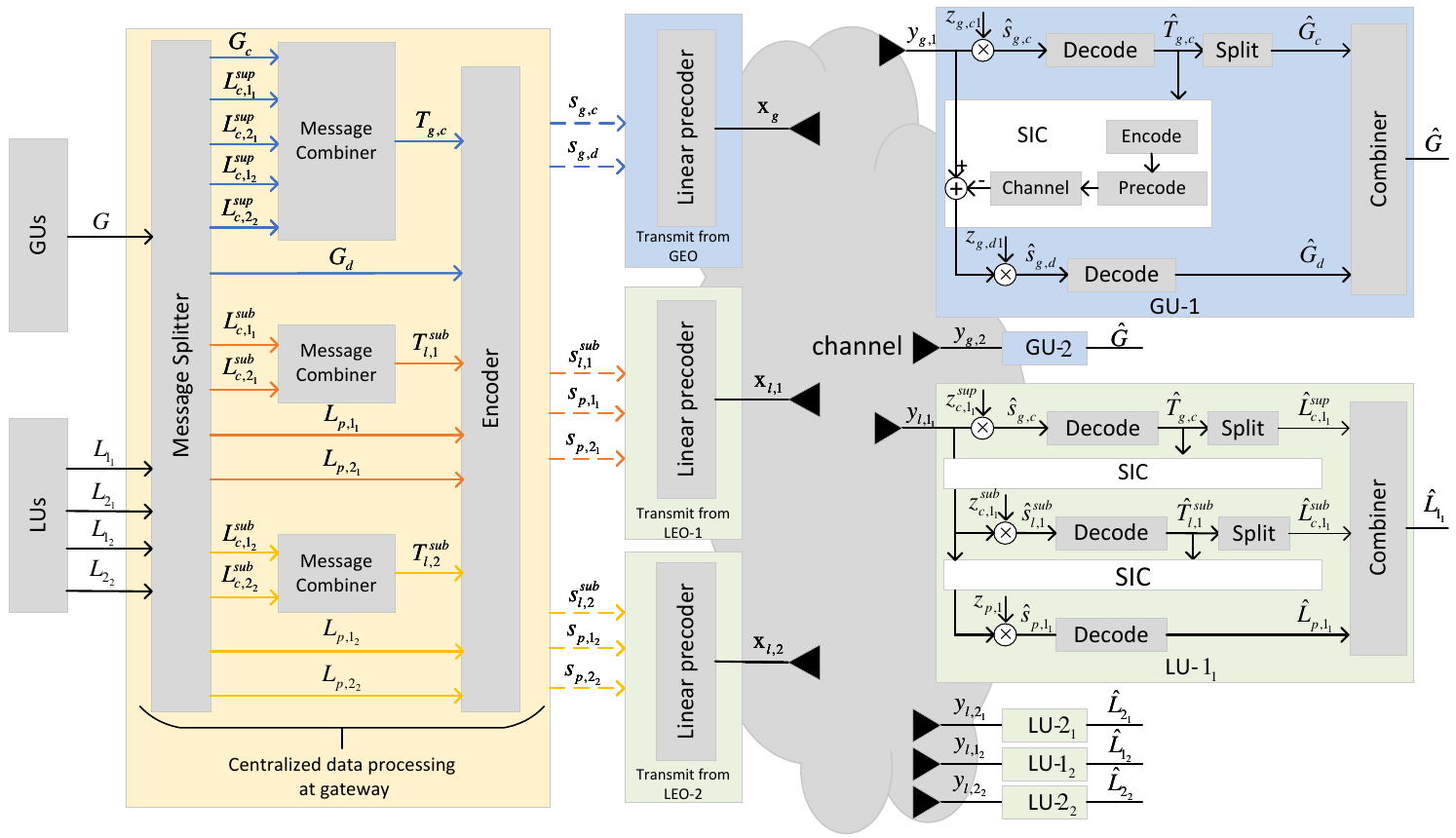}
	\caption{\gls{drsma} signal transmission model for 1 \gls{geo} and 2 \gls{leo} satellites deployment, $N=2$, $M=4$, each \gls{leo} satellite serves $2$ \gls{lus}.\vspace{-5mm}}
	\label{fig:sig_illu}
\end{figure*}

We assume that all \gls{gus} are interested in the same content, and the multicast message $G$ is split into a common part, $G_{c}$, and a \gls{gus}-designated part, $G_{d}$, in \gls{drsma} scheme.
The unicast messages $L_1,\ldots,L_{K_m}$ for \gls{lus} served by the $m$-th \gls{leo} are split into three parts, namely, a super-common part, a sub-common part and a private part, i.e., $L_{k_m}\rightarrow\left\{L_{c,k_m}^{sup},\ L_{c,k_m}^{sub},\ L_{p,k_m}\right\}$, $\forall k_m\in\mathcal{K}_m,\ m\in\mathcal{M}$. The superscript ``\textit{sup}'' and ``\textit{sub}'' are used to represent the super-common and sub-common parts for \gls{lus} messages/streams, respectively.
The common message for \gls{gus}, $G_{c}$, and all super-common messages for \gls{lus}, $\{\bigcup_{k_m\in\mathcal{K}}L_{c,k}^{sup}\}$, are combined as $T_{g,c}$ and encoded into $s_{g,c}$ to be transmitted from \gls{geo} and decoded by all \gls{gus} and \gls{lus}. It manages the inter-system interference between \gls{geo} and \gls{leo} so as the inter-\gls{leo} interference. The \gls{gus}-designated message, $G_{d}$, is encoded into $s_{g,d}$. The vector of \gls{geo} streams is $\mathbf{s}_g={\left[s_{g,c},s_{g,d}\right]}^T$ and $\mathbb{E}\left\{\mathbf{s}_g{\mathbf{s}_g}^H\right\}=\mathbf{I}$. The sub-common messages, $\{\bigcup_{k_m\in\mathcal{K}_m}L_{c,k_m}^{sub}\}$, for \gls{lus} under $m$-th \gls{leo} coverage are combined into a sub-common message $T_{l,m}^{sub}$ and encoded into a sub-common stream $s_{l,m}^{sub}$ to manage the intra-\gls{leo} interference.
The subscript ``\textit{g}'' and ``\textit{l}'' are used to represent messages/streams from the \gls{geo} and \gls{leo} satellites, respectively.
The private parts for \gls{lus} are independently encoded into private streams $s_{p,1},\ldots,s_{p,K_m}$. The vector of $m$-th \gls{leo} satellite streams $\mathbf{s}_{l,m}={\left[s_{l,m}^{sub},s_{p,1},\ldots,s_{p,K_m}\right]}^T\in\mathbb{C}^{(K_m+1)\times1}$ is obtained, and we assume it satisfies $\mathbb{E}\left\{\mathbf{s}_{l,m}\mathbf{s}_{l,m}^H\right\}=\mathbf{I}$.
A signal transmission model for \gls{drsma} is shown in Fig. \ref{fig:sig_illu}.
\par Data streams are mapped to the transmit antennas via precoding matrix $\mathbf{W}=\left[\mathbf{w}_{c},\mathbf{w}_{d}\right]\in\mathbb{C}^{N_{tg}\times2}$ at \gls{geo} satellite and precoding matrices $\mathbf{P}_{m}=[\mathbf{p}_{c,m},\mathbf{p}_{p,1},\ldots,\mathbf{p}_{p,K_m}]$ at $m$-th \gls{leo} satellite, where $\mathbf{P}_{m}\in\mathbb{C}^{N_{tl}\times(K_m+1)}$. The respective transmit signals at the \gls{geo} and $m$-th \gls{leo} are
\vspace{-2mm}
\begin{subequations}\label{equ:geo_trans_signal}
	\begin{align}
		 & \mathbf{x}_g=\mathbf{w}_{c}s_{g,c}+\mathbf{w}_{d}s_{g,d},                                                       \\
		 & \mathbf{x}_{l,m}=\mathbf{p}_{c,m}s_{l,m}^{sub}+\sum_{k_m=1}^{K_m}\mathbf{p}_{p,k_m}s_{p,k_m},\ m\in\mathcal{M}.
	\end{align}
\end{subequations}
\vspace{-1mm}
Following \cite{6184256,9628071}, we assume that they are subject to the total power constraints, i.e., $\mathbb{E}\left\{\mathbf{x}_{g}^H\mathbf{x}_{g}\right\}\leq P_{g}$, and $\mathbb{E}\left\{\mathbf{x}_{l,m}^H\mathbf{x}_{l,m}\right\}\leq P_{l}$.
The received signal at \gls{gus} and \gls{lus} are illustrated in the following two subsections, respectively.

\subsubsection{GEO UT received signal}
After each \gls{gu} receives the signal, it first decodes the \gls{geo} common stream, $s_{g,c}$, by treating other streams as noise. The interference between \gls{geo} and \gls{leo} satellites and inter-\gls{leo} interference are handled by $s_{g,c}$, because it contains a portion of UT messages from both \gls{gus} and \gls{lus}. Each \gls{gu} or \gls{lu} partially decodes the inter-system interference and partially treats inter-system interference as noise, so as inter-\gls{leo} interference for \gls{lus}.
The signal received at \gls{gu}-$n$ can be expressed as
\vspace{-2mm}
\begin{equation}\label{equ:geo_ut_rece_sig}
	\begin{aligned}
		y_{g,n}  =\  & \mathbf{h}^{H}_{g,n} \mathbf{x}_{g}+\sum_{m=1}^M \mathbf{h}_{l2g,mn}^H\mathbf{x}_{l,m} + n_n                                                                                  \\
		=\           & \underbrace{\mathbf{h}^{H}_{g,n}\mathbf{w}_{c}s_{g,c}+\mathbf{h}^{H}_{g,n}\mathbf{w}_{d}s_{g,d}}_{\text{Desired signal}}                                                      \\
		             & +\underbrace{\sum^{M}_{m=1}\mathbf{h}^H_{l2g,mn}(\mathbf{p}_{c,m}s_{l,m}^{sub}+\sum_{i=1}^{K_m}\mathbf{p}_{p,i}s_{p,i})}_{\text{Interference from \gls{leo} satellites}}+n_n,
	\end{aligned}
\end{equation}
where $\mathbf{h}_{g,n}\in\mathbb{C}^{N_{tg}\times1}$ is the channel between the \gls{geo} satellite and \gls{gu}-$n$ 
and $\mathbf{h}_{l2g,mn}\in\mathbb{C}^{N_{tl}\times1}$ is the channel between the $m$-th \gls{leo} satellite and \gls{gu}-$n$, $m\in\mathcal{M}$. $n_n\sim\mathcal{CN}\left(0,\sigma^2_n\right)$ represents the \gls{awgn} at \gls{gu}-$n$, $n\in\mathcal{N}$. 

\subsubsection{LEO UT received signal}
\par The received signal at \gls{lu}-$k_m$ writes as
\begin{equation}\label{equ:received_sig}
	\hspace{-2mm}
	\begin{aligned}
		y_{l,k_m}  =\  & \mathbf{h}^{H}_{l,mk_m} \mathbf{x}_{l,m}+\mathbf{h}^{H}_{g2l,k_m} \mathbf{x}_{g} + \sum_{\substack{j=1,                                                                                  \\j\neq m}}^M \mathbf{h}_{l,jk_m}^H\mathbf{x}_{l,j}+n_{k_m}                                                                 \\
		=\             & \underbrace{\mathbf{h}^{H}_{g2l,k_m}\mathbf{w}_{c}s_{g,c}+\mathbf{h}^{H}_{l,mk_m}\mathbf{p}_{c,m}s_{l,m}^{sub}+\mathbf{h}^H_{l,mk_m}\mathbf{p}_{p,k_m}s_{p,k_m}}_{\text{Desired signal}} \\
		               & +\underbrace{\mathbf{h}^H_{l,mk_m}\sum^{K_m}_{\substack{i=1,                                                                                                                             \\i\neq k_m}}\mathbf{p}_{p,i}s_{p,i}}_{\text{Intra-\gls{leo} interference}} +\underbrace{\mathbf{h}^{H}_{g2l,k_m}\mathbf{w}_{d}s_{g,d}}_{\text{Interference from \gls{geo} satellite}}\\
		               & +\underbrace{\sum^M_{\substack{j=1,                                                                                                                                                      \\j\neq m}}\mathbf{h}^{H}_{l,jk_m}\left(\mathbf{p}_{c,j}s_{l,j}^{sub}+\sum^{K_j}_{i=1}\mathbf{p}_{p,i}{s}_{p,i}\right)}_{\text{Inter-\gls{leo} interference}}+n_{k_m},
	\end{aligned}
\end{equation}
where $\mathbf{h}_{l,mk_m}\in\mathbb{C}^{N_{tl}\times1}$ is the channel between the $m$-th \gls{leo} satellite and \gls{lu}-$k_m$ 
and $\mathbf{h}_{g2l,k_m}\in\mathbb{C}^{N_{tg}\times1}$ is the channel between the \gls{geo} satellite and \gls{lu}-$k_m$. $n_{k_m}\sim\mathcal{CN}\left(0,\sigma^2_{k_m}\right)$ represents the \gls{awgn} at \gls{lu}-$k_m$, $k_m\in\mathcal{K}_m$, $m\in\mathcal{M}$. 

\subsection{Channel Model}\label{sec:channel_model}
In this subsection, we illustrate the channel models of \gls{geo} and \gls{leo} satellites.
\subsubsection{GEO Satellite Channel}
We assume that the channel realizations between the satellite and different \gls{uts} are uncorrelated, because different \gls{uts} are often geographically separated by a certain distance.
The main characteristics of \gls{geo} channel include atmospheric fading and radiation pattern. Rain attenuation is the dominant atmospheric impact for the Ka-band satellite channel, and it is often characterized by a lognormal distribution \cite{9257433,9844445}.
The \gls{geo} satellite channel vector between the satellite and $i$-th UT is expressed as $\mathbf{h}_{g,i}=[h_{g,1i},h_{g,2i},\ldots,h_{g,N_{tg} i}]^T$, $i\in\mathcal{N}\cup\mathcal{K}$. The $n_{tg}$-th entry of the channel between the \gls{geo} satellite and the UT-$i$ can be written as
\begin{equation}\label{equ:GEO_channel}
	\begin{aligned}
		h_{g,n_{tg}i}=\frac{\sqrt{G_{r}G_{g,i}}}{4\pi\frac{d_{g,i}}{\lambda}\sqrt{\kappa T_{sys}B_w}}\chi_{n_{tg}i}^{-\frac{1}{2}}e^{-j{\phi}_{g,n_{tg}i}},i\in\left\{\mathcal{N}\cup\mathcal{K}\right\},
	\end{aligned}
\end{equation}
where $G_{r}$ and $G_{g,i}$ are the UT antenna gain and the antenna gain from \gls{geo} satellite to the UT-$i$, respectively. $d_{g,i}$ and $\lambda$ denote the distance from the \gls{geo} satellite to UT-$i$ and the carrier wavelength, respectively. $\kappa$ is the Boltzmann constant, $T_{sys}$ is the receiving system noise temperature and $B_w$ denotes the user link bandwidth. The large-scale fading is modeled as $\chi_{n_{tg}i,dB}=10\log_{10}(\chi_{n_{tg}i})$, and $\ln(\chi_{n_{tg}i,dB})\sim\mathcal{N}(\mu,\sigma)$. $\boldsymbol{\phi}_{g,i}=[\phi_{g,1i},\phi_{g,2i},\ldots,\phi_{N_{tg}i}]^T\in\mathbb{R}^{N_{tg}\times1}$ is channel phase vector. 
$G_{g,i}$ can be further expressed as
\begin{equation}\label{equ:GEO_channel_sa_gain}
	G_{g,i}=G_{max}{\left[\frac{J_1(u_{g,i})}{2u_{g,i}}+36\frac{J_3(u_{g,i})}{u_{g,i}^3}\right]}^2,
\end{equation}
where $u_{g,i}=2.07123\frac{\sin(\theta_{g,i})}{\sin(\theta_{3dB})}$. $\theta_{g,i}$ is the angle between UT-$i$ and the center of \gls{geo} satellite beam. $\theta_{3dB}$ denotes a $3$ dB loss angle compared with the beam center. $G_{max}$ is the maximum beam gain of the beam center. $J_1$ and $J_3$ are first-kind Bessel functions with first-order and third-order, respectively. The channel between \gls{geo} and \gls{lu}-$k_m$ is $\mathbf{h}_{g2l,k_m}=[h_{g,1k_m},h_{g,2k_m},\ldots,h_{g,N_{tg} k_m}]^T$, $k_m\in\mathcal{K}_m, m\in\mathcal{M}$. 

\subsubsection{LEO Satellite Channel}
Due to the high mobility of \gls{leo}, the propagation characteristics of satellite channel and impact on the channel modelling are different. It differs from \gls{geo} satellite channel due to the Doppler shift and delay \cite{9110855,9628071}.
The Doppler shift $f_{i,u}$ associated with propagation path $u$ of UT-$i$ is dominated by two independent Doppler shifts, $f_{i,u}^{sat}$ and $f_{i,u}^{ut}$, which result from the movements of \gls{leo} satellite and UT, respectively. The channel between $m$-th \gls{leo} satellite and $i$-th UT is
\begin{subequations}\label{equ:LEO_channel}
	\begin{align}
		 & \mathbf{h}_{l,mi}=g_{mi}(t,f)e^{j2\pi(f_{i,u}^{sat}t-f\tau_{i}^{\min})}e^{-j\boldsymbol{\phi}_{l,mi}},                          \\
		 & g_{mi}(t,f)=\sum_{u=0}^{U_i-1}g_{i,u}e^{j2\pi(f_{i,u}^{ut}t-f\tau_{i,u}^{ut})},\ i\in\left\{\mathcal{N}\cup\mathcal{K}\right\},
	\end{align}
\end{subequations}
where $\tau_{i}^{\min}$ is the minimum value of the propagation delays of the $i$-th UT defined by $\tau_{i}^{\min}=\min_u\{\tau_{i,u}\}$, where $\tau_{i,u}$ is the delay of the $u$-th path to UT-$i$. We denote the delay difference of UT-$i$ with path $u$ as $\tau_{i,u}^{ut}\triangleq\tau_{i,u}-\tau_i^{\min}$. $U_i$ denotes the number of propagation paths of the $i$-th UT. The Doppler shift $f_{i,u}^{sat}$ induced by the movement of the \gls{leo} satellite can be considered to be identical for different propagation paths of the same UT-$i$ due to the relatively high altitude of the \gls{leo} satellite \cite{966585,9110855}. Hence, in order to simplify the notation, we recast and omit the path index the Doppler shifts caused by the motion of the \gls{leo} satellite as $f_{i,u}^{sat}=f_{i}^{sat}$. $\boldsymbol{\phi}_{l,mi}$ is the channel phase vector from \gls{leo} satellite-$m$ to UT-$i$, $\boldsymbol{\phi}_{l,mi}=[\phi^{l,m}_{1i},\phi^{l,m}_{2i},\ldots,\phi^{l,m}_{N_{tl}i}]^T\in\mathbb{R}^{N_{tl}\times1}$, and it can be calculated by using geographic location information between the $i$-th UT and the \gls{leo} satellite, or it can be acquired through using satellite positioning system \cite{Bagrov2015}. $g_{mi}(t,f)$ is \gls{leo} satellite downlink channel gain, $g_{i,u}$ is the complex-valued gain corresponding to path $u$ and UT-$i$ \cite{9852292}. \gls{leo} satellite system usually operates under \gls{los} propagations, and we assume $g_{mi}(t,f)$ follows Rician fading distribution with the Rician factor $\kappa_i$ and power
\begin{equation}\label{equ:LEO_channel_sa_gain}
	\begin{aligned}
		\gamma_i=\mathbb{E}\left\{{|g_{mi}(t,f)|}^2\right\}=\frac{G_{r}G_{l}}{(4\pi\frac{d_{mi}}{\lambda})^2{\kappa T_{sys}B_w}}
	\end{aligned},
\end{equation}
where $G_{l}$ is the antenna gain of the \gls{leo} satellite. Alternatively, the real and imaginary parts of $g_{mi}(t,f)$ have \gls{iid} real-valued Gaussian entries with a certain mean and variance, i.e., $g_{mi}(t,f)\sim\mathcal{N}\left(\sqrt{\frac{\kappa_i\gamma_i}{2(\kappa_i+1)}},\frac{\gamma_i}{2(\kappa_i+1)}\right)$.
Following \cite{9149121,9110855}, we assume that the time and frequency compensation can be properly performed at \gls{uts}. Specifically, the Doppler compensation, $f_{i}^{syn}=f_{i}^{sat}$, and delay compensation, $\tau_i^{syn}=\tau_i^{\min}$ are applied. 

\subsection{Imperfect channel}
\par Due to the long propagation delay between a satellite and \gls{uts} as well as the mobility of satellites and \gls{uts}, it is difficult to obtain instantaneous \gls{csi}. Hence, we assume that statistical \gls{csit} and \gls{csir} are available in the network. 
\par Under imperfect \gls{csit} scenario, we utilize ergodic rate as a long-term measurement to capture the expected performance over a known statistical channel distribution. With imperfect \gls{csir} deployment, the progress of \gls{sic} is effected and the stream cannot be perfectly removed, since \gls{uts} lack of accurate channel information.
We denote \gls{geo} channel phase vector $\boldsymbol{\phi}_{g,i}$ as the sum of a estimated phase, $\hat{\boldsymbol{\phi}}_{g,i}$, and estimation error, ${\boldsymbol{\phi}}_{eg,i}$, i.e., $\boldsymbol{\phi}_{g,i}=\hat{\boldsymbol{\phi}}_{g,i}+{\boldsymbol{\phi}}_{eg,i}$, $i\in\mathcal{N}\cup\mathcal{K}$ \cite{9815078}.
The channel phase is influenced by multiple time-varying factors, such as the rain, cloud, gaseous absorption and the use of different Local Oscillators (LO) onboard the satellite for each antenna feed \cite{7765141}.
	The phase estimation error, ${\boldsymbol{\phi}}_{eg,i} \sim \mathcal{N}(\mathbf{0},\sigma_{e,g}^2\mathbf{I})$, is characterized by a normal distribution with a variance of $\sigma_{e,g}^2$ that is affected by the types of LO.
The relationship between the actual channel and estimated channel can be expressed as $\mathbf{h}_{g,i}=\hat{\mathbf{h}}_{g,i}\circ\boldsymbol{\Phi}_{eg,i}$, where $\boldsymbol{\Phi}_{eg,i} = \exp(-j\boldsymbol{\phi}_{eg,i})$. 
Similarly, the $m$-th \gls{leo} channel phase vector can be written as $\boldsymbol{\phi}_{l,mi}=\hat{\boldsymbol{\phi}}_{l,mi}+\boldsymbol{\phi}_{el,mi}$, where ${\boldsymbol{\phi}}_{el,mi}\sim\mathcal{N}(\mathbf{0},\sigma_{e,l}^{2}\mathbf{I})$. The channel vector from $m$-th \gls{leo} satellite to $i$-th UT is $\mathbf{h}_{l,mi}=\hat{\mathbf{h}}_{l,mi}\circ\boldsymbol{\Phi}_{el,mi}$, where $\boldsymbol{\Phi}_{el,mi} = \exp(-j\boldsymbol{\phi}_{el,mi})$. 
$\boldsymbol{\phi}_j=\mathbf{0}$, $j\in\{\{eg,i\},\{el,mi\}\}$, represents prefect \gls{csi}.

\subsection{Signal transmission and rate expressions}
In this subsection, the signal transmission progress, \gls{sinr} and ergodic rate expressions are illustrated.
\subsubsection{GEO UT}
Once receiving the signal, \gls{gu}-$n$ first decodes the common stream $s_{g,c}$. Under imperfect \gls{csir}, 
\gls{uts} only have statistical \gls{csi}, therefore, \gls{uts} decode the desired stream based on the estimated channel \cite{an2021rate}. The \gls{sinr} of decoding the common stream at \gls{gu}-$n$ is expressed as

\begin{equation}\label{equ:geo_com_snir}
	\rho_{g,cn}=  \frac{\hat{S}_{g,cn}}{\tilde{S}_{g,cn}+ S_{g,dn}+I_{g,n}^{tot}+\sigma_n^2},\ n\in\mathcal{N},
\end{equation}
where $\hat{S}_{g,cn}=|\hat{\mathbf{h}}^{H}_{g,n}\mathbf{w}_{c}|^2$ and
$S_{g,dn}=|\mathbf{h}^H_{g,n}\mathbf{w}_{d}|^2$.
$\tilde{S}_{g,cn}=|{\mathbf{h}}^{H}_{g,n}\mathbf{w}_{c}|^2-|\hat{\mathbf{h}}^{H}_{g,n}\mathbf{w}_{c}|^2$ is residual power.
$I_{g,n}^{tot}=\sum^{M}_{m=1}|\mathbf{h}^H_{l2g,mn}\mathbf{p}_{c,m}|^2+\sum^M_{m=1}\sum^{K_m}_{i=1}|\mathbf{h}^H_{l2g,mn}\mathbf{p}_{p,i}|^2$ is the sum of interference power of sub-common streams and the private streams for \gls{lus}.
The corresponding ergodic rate expresses as $R_{g,cn}=\mathbb{E}\{\log_2(1+\rho_{g,cn})\}$. 
The received signal after decoding $s_{g,c}$ writes as
\begin{equation}\label{equ:sig_after_sic}
	\begin{aligned}
		y_{g,n}^{SIC} =\  & y_{g,n} - \hat{\mathbf{h}}^{H}_{g,n}\mathbf{w}_{c}                                                                                        \\
		=\                & \hat{\mathbf{h}}^{H}_{g,n}(\diag{(\boldsymbol{\Phi}_{eg,n}^H)}-\mathbf{I})\mathbf{w}_{c}s_{g,c}+\mathbf{h}^{H}_{g,n}\mathbf{w}_{d}s_{g,d} \\
		\                 & +\sum^{M}_{m=1}\mathbf{h}^H_{l2g,mn}(\mathbf{p}_{c,m}s_{l,m}^{sub}+\sum_{i=1}^{K_m}\mathbf{p}_{p,i}s_{p,i})+n_n                           \\
		=\                & \tilde{\mathbf{h}}^{H}_{g,n}\mathbf{w}_{c}s_{g,c}+\mathbf{h}^{H}_{g,n}\mathbf{w}_{d}s_{g,d}                                               \\
		\                 & +\sum^{M}_{m=1}\mathbf{h}^H_{l2g,mn}(\mathbf{p}_{c,m}s_{l,m}^{sub}+\sum_{i=1}^{K_m}\mathbf{p}_{p,i}s_{p,i})+n_n,
	\end{aligned}
\end{equation}
where $\tilde{\mathbf{h}}^{H}_{g,n}$ is denoted as residual channel in this paper. Hence, $\tilde{S}_{g,cn}$ can be rewritten as $\tilde{S}_{g,cn}=|\tilde{\mathbf{h}}^{H}_{g,n}\mathbf{w}_{c}|^2$.

\par Subsequently, \gls{gu}-$n$ decodes the \gls{gus}-designated stream $s_{g,d}$. Accordingly, \gls{gu}-$n$'s decoding \gls{sinr} of the \gls{gus}-designated stream writes as
\begin{equation}\label{equ:geo_priv_sinr}
	\rho_{g,dn}=\frac{\hat{S}_{g,dn}}{\tilde{S}_{g,cn}+\tilde{S}_{g,dn}+I_{g,n}^{tot}+\sigma_n^2},\ n\in\mathcal{N},
\end{equation}
where $\tilde{S}_{g,dn}$ is the residual power of decoding $s_{g,d}$, and $\tilde{S}_{g,dn}=|\tilde{\mathbf{h}}^{H}_{g,n}\mathbf{w}_{d}|^2$.
The ergodic designated rate of \gls{gu}-$n$ is expressed as $R_{g,dn}=\mathbb{E}\{\log_2(1+\rho_{g,dn})\}$. \gls{gu}-$n$ reconstructs the original message after the common and \gls{gus}-designated messages have been decoded by taking the decoded ${G}_{c}$ from the decoded ${T}_{g,c}$ and combining it with the decoded $G_{d}$.

\subsubsection{LEO UT}
Based on estimated channel, \gls{lu}-$k_m$ firstly decodes the $s_{g,c}$ by treating other interference as noise. The corresponding \gls{sinr} is
\vspace{-2mm}
\begin{equation}\label{equ:leo_com_sinr}
	\rho_{l,k_m}^{sup} =  \frac{\hat{S}_{g2l,ck_m}}{\tilde{S}_{g2l,ck_m}+S_{l,ck_m}+S_{l,pk_m}+I_{l,k_m}^{tot}+\sigma_{k_m}^2},\ k_m\in\mathcal{K}_m,
\end{equation}
where $\hat{S}_{g2l,ck_m}=|\hat{\mathbf{h}}^{H}_{g2l,k_m}\mathbf{w}_{c}|^2$,
$\tilde{S}_{g2l,ck_m} = |\tilde{\mathbf{h}}^{H}_{g2l,k_m}\mathbf{w}_{c}|^2$,
$S_{l,ck_m}=|\mathbf{h}^H_{l,mk_m}\mathbf{p}_{c,m}|^2$,
$S_{l,pk_m}=|\mathbf{h}^H_{l,mk_m}\mathbf{p}_{p,k_m}|^2$. $\tilde{\mathbf{h}}^{H}_{g2l,k_m}=\hat{\mathbf{h}}^{H}_{g2l,k_m}(\diag{(\boldsymbol{\Phi}_{eg,k_m}^H)}-\mathbf{I})$ denotes the residual channel between \gls{geo} satellite and \gls{lu}-$k_m$.
$I_{l,k_m}^{tot}=\allowbreak\sum_{\substack{i=1,\\i\neq k_m}}^{K_m}|\mathbf{h}^H_{l,mk_m}\mathbf{p}_{p,i}|^2\allowbreak+\sum^{M}_{\substack{j=1,\\j\neq m}}|\mathbf{h}^H_{l,jk_m}\mathbf{p}_{c,j}|^2\allowbreak+\sum^{M}_{\substack{j=1,\\j\neq m}}\sum^{K_j}_{i=1}|\mathbf{h}^H_{l,jk_m}\mathbf{p}_{p,i}|^2+|\mathbf{h}^{H}_{g2l,k_m}\mathbf{w}_{d}|^2$ is the sum of interference power received at \gls{lu}-$k_m$.
The corresponding ergodic rate is $R_{l,k_m}^{sup}=\mathbb{E}\{\log_2(1+\rho_{l,k_m}^{sup})\}$.
In order to make sure that all \gls{gus} and \gls{lus} can decode $s_{g,c}$, we define the common rate
\vspace{-2mm}
\begin{equation}\label{equ:com_rate_expre}
	R_c=\min_{n\in\mathcal{N},k\in\mathcal{K}}\ \left\{R_{g,cn},\ R_{l,k}^{sup}\right\}.
\end{equation}
Since $s_{g,c}$ is shared among all \gls{gus} and \gls{lus}, we define $R_c\triangleq\sum_{n=1}^{N}C_{g,n}+\sum_{m=1}^{M}\sum_{k_m=1}^{K_m}C_{l,k_m}^{sup}$, where $C_{g,n}$ and $C_{l,k_m}^{sup}$ denote the $n$-th \gls{gu}'s and $k_m$-th \gls{lu}'s portions of common rate, respectively.
\par Once $s_{g,c}$ is decoded, its contribution to the original received signal $y_{l,k_m}$ is subtracted through \gls{sic}. After that, \gls{lu}-$k_m$ decodes its sub-common stream $s_{l,m}^{sub}$ by treating other sub-common streams as noise. The sub-common streams are used to manage intra-\gls{leo} interference since it contains parts of messages for \gls{lus} served by the same \gls{leo}. Sub-common streams enable users to partially decode intra-\gls{leo} interference and partially treat intra-\gls{leo} interference as noise. The \gls{sinr} of decoding the sub-common stream $s_{l,m}^{sub}$ at \gls{lu}-$k_m$ served by \gls{leo}-$m$ is
\vspace{-2mm}
\begin{equation}\label{equ:leo_sub_com_sinr}
	\rho_{l,k_m}^{sub}= \frac{\hat{S}_{l,ck_m}}{\tilde{S}_{g2l,ck_m} +\tilde{S}_{l,ck_m} + S_{l,pk_m}+I_{l,k_m}^{tot}+\sigma_{k_m}^2},\ k_m\in\mathcal{K}_m,
\end{equation}
where $\tilde{S}_{l,ck_m} = |\tilde{\mathbf{h}}^{H}_{l,mk_m}\mathbf{p}_{c,m}|^2$. $\tilde{\mathbf{h}}^{H}_{l,mk_m}=\hat{\mathbf{h}}^{H}_{l,mk_m}(\diag{(\boldsymbol{\Phi}_{el,mk_m}^H)}-\mathbf{I})$ is the residual channel between \gls{leo} satellite $m$ and \gls{lu}-$k_m$.
The ergodic rate of the sub-common stream is $R_{l,k_m}^{sub}=\mathbb{E}\{\log_2(1+\rho_{l,k_m}^{sub})\}$. To guarantee that all \gls{lus} served by $m$-th \gls{leo} are capable of decoding $s_{l,m}^{sub}$, we define the achievable sub-common rates as
\vspace{-2mm}
\begin{equation}\label{equ:leo_sub_com_rate_expre}
	R_{c,m}^{sub}=\min_{k_m\in\mathcal{K}_m}\ \left\{R_{l,k_m}^{sub}\right\}=\sum_{k_m=1}^{K_m}C_{l,k_m}^{sub},
\end{equation}
where $C_{l,k_m}^{sub}$ is the rate at which $L_{c,k_m}^{sub}$ is communicated. After the sub-common stream is decoded, re-encoded, precoded, and subtracted from the received signal through \gls{sic}, \gls{lu}-$k_m$ decodes its private stream $s_{k_m}$.
The \gls{sinr} of decoding the private stream $s_{k_m}$ at \gls{lu}-$k_m$ is
\vspace{-2mm}
\begin{equation}\label{equ:leo_priv_sinr}
	\rho_{k_m}= \frac{\hat{S}_{l,pk_m}}{\tilde{S}_{g2l,ck_m} +\tilde{S}_{l,ck_m}+ \tilde{S}_{l,pk_m} +I_{l,k_m}^{tot}+\sigma_{k_m}^2},\ k_m\in\mathcal{K}_m,
\end{equation}
where $\tilde{S}_{l,pk_m} = |\tilde{\mathbf{h}}^{H}_{l,mk_m}\mathbf{p}_{p,k_m}|^2$.
The corresponding ergodic private rate is $R_{p,k_m}=\mathbb{E}\{\log_2(1+\rho_{k_m})\}$. Thus, the achievable rates of \gls{gu}-$n$ and \gls{lu}-$k_m$ are respectively given as
\begin{subequations}\label{equ:ut_rates}
	\begin{align}
		 & R_n=C_{g,n}+R_{g,dn},\ n\in\mathcal{N},                                                    \\
		 & R_{k_m}=C_{l,k_m}^{sup}+C_{l,k_m}^{sub}+R_{p,k_m},\ k_m\in\mathcal{K}_m,\ m\in\mathcal{M}.
	\end{align}
\end{subequations}
\begin{remark}
	\gls{drsma} is designed for managing interference of the co-existence \gls{geo}/\gls{leo} satellites multilayer network\footnote{This paper primarily focuses on the multilayer satellites network. However, it is	worth noting that the proposed \gls{drsma} can also be extended and applied to heterogeneous wireless communication systems, e.g., multilayer \gls{uav} network or terrestrial heterogeneous wireless communications.}. Each \gls{gu} decodes $s_{g,c}$ first and requires $1$ layer of \gls{sic} to decode $s_{g,d}$. By decoding $s_{g,c}$, the inter-system interference can be suppressed since partial inter-system interference is decoded and part of inter-system interference is treated as noise. While each \gls{lu} requires $2$ layers of \gls{sic} before decoding intended private stream (in order to decode $s_{g,c}$ first and then decode $s_{l,m}^{sub}$). The inter-system interference and inter-\gls{leo} interference is managed through $s_{g,c}$, and the intra-\gls{leo} interference is managed by $s_{l,m}^{sub}$.
	\par \gls{drsma} is a general framework of multilayer network that encompasses traditional \gls{rsma} and \gls{sdma} as special cases. By switching off $s_{g,c}$, the multiple access technique of \gls{geo} sub-network becomes traditional multicasting, and the multiple access approach of \gls{leo} sub-network reduces to 1-layer \gls{rsma}. By turning off $s_{g,c}$ and sub-common streams, \gls{geo} and \gls{leo} sub-networks work in multicasting and \gls{sdma} manner, respectively.
\end{remark}

\section{Problem Formulation and Proposed Algorithm}\label{sec:Prob_formu}
\par In this section, we focus on designing a joint optimization of \gls{geo}-\gls{leo} beamforming problem that maximizes the minimum fairness rate across all \gls{gus} and \gls{lus} under the constraint of transmit power at all satellites. 
For \gls{drsma} based multilayer satellite network, the optimization problem can be formulated as
\begin{subequations}\label{prob:formu_prob_orig}
	\begin{align}
		\mathcal{P}_{1}: & \max_{\mathbf{c},\mathbf{W},\mathbf{P}}
		\ \min_{n\in\mathcal{N},k\in\mathcal{K}}\quad  \left\{R_n,R_k\right\}\label{prob:formu_prob_orig_a}                                      \\
		\mbox{s.t.}      & \quad\quad\sum_{n\in\mathcal{N}}C_{g,n}+\sum_{k\in\mathcal{K}}C_{l,k_m}^{sup}\leq R_{c}\label{prob:formu_prob_orig_b} \\
		                 & \quad\quad\sum_{k_m\in\mathcal{K}_m}C_{l,k_m}^{sub}\leq R_{c,m}^{sub},\ m\in\mathcal{M}\label{prob:formu_prob_orig_c} \\
		                 & \quad\quad\mathrm{tr}(\mathbf{W}{\mathbf{W}}^{H})\leq P_g\label{prob:formu_prob_orig_d}                               \\
		                 & \quad\quad\mathrm{tr}(\mathbf{P}_{m}\mathbf{P}_{m}^{H})
		\leq P_l,\ m\in\mathcal{M}\label{prob:formu_prob_orig_e}                                                                                 \\
		                 & \quad\quad\mathbf{c}\succeq\mathbf{0},\label{prob:formu_prob_orig_f}
	\end{align}
\end{subequations}
where $\mathbf{P}=[\mathbf{P}_{1},\ldots,\mathbf{P}_{M}]$ represents the precoding matrix at all \gls{leo} satellites. $\mathbf{c}=[\mathbf{c}_{g},\mathbf{c}_{l}^{sup},\mathbf{c}_{l}^{sub}]$, where $\mathbf{c}_{g}=[C_{g,1},\ldots,C_{g,N}]^T$, $\mathbf{c}_{l}^{sup}=[C_{l,1}^{sup},\ldots,C_{l,K}^{sup}]^T$, and $\mathbf{c}_{l}^{sub}=[C_{l,1}^{sub},\ldots,C_{l,K}^{sub}]^T$ are the vectors of common rate portions. (\ref{prob:formu_prob_orig_b}) ensures that all \gls{gus} and \gls{lus} can successfully decode the common stream $s_{g,c}$. Similarly, (\ref{prob:formu_prob_orig_c}) guarantees that the sub-common stream $s_{l,m}^{sub}$ can be decoded by all \gls{lus} under the coverage of \gls{leo}-$m$. (\ref{prob:formu_prob_orig_d}) and (\ref{prob:formu_prob_orig_e}) represent power constraints of the \gls{geo} and \gls{leo} satellites. (\ref{prob:formu_prob_orig_f}) ensure that all common rate portions are non-negative.

The formulated problem (\ref{prob:formu_prob_orig}) is nonconvex, while it can be transformed into a tractable \gls{sdp} problem and solved with rank-one constraints. We propose a robust \gls{sdp}-based iterative optimization algorithm to solve the problem with statistical \gls{csit} and \gls{csir} in this work.
Denote $\boldsymbol{\alpha}=\left\{\boldsymbol{\alpha}_{g,c},\ \boldsymbol{\alpha}_{g,p},\ \allowbreak\boldsymbol{\alpha}_{l}^{sup},\ \allowbreak\boldsymbol{\alpha}_{l}^{sub},\ \allowbreak\boldsymbol{\alpha}_{p}\right\}$ and $\mathbf{F}=\allowbreak\big\{\mathbf{F}_{g,c},\ \allowbreak\mathbf{F}_{g,d},\ \allowbreak\mathbf{F}_{l,cm},\ \allowbreak\mathbf{F}_{l,pk}|k\in\mathcal{K},\allowbreak\ m\in\mathcal{M}\big\}$, where
$\mathbf{F}_{g,c}=\mathbf{w}_{c}\mathbf{w}_{c}^H$, $\mathbf{F}_{g,d}=\mathbf{w}_{d}\mathbf{w}_{d}^H$, ${\{\mathbf{F}_{l,cm}=\mathbf{p}_{c,m}\mathbf{p}_{c,m}^H|m\in\mathcal{M}\}}$ and ${\{\mathbf{F}_{l,pk}=\mathbf{p}_{l,pk}\mathbf{p}_{l,pk}^H|k\in\mathcal{K}\}}$.
By introducing auxiliary variables $t$, $\boldsymbol{\alpha}$ and $\mathbf{F}$, the equivalent reformulation of $\mathcal{P}_1$ is
\begin{subequations}\label{prob2:formu_prob_sdp}
	\hspace{-6mm}
	\begin{align}
		\mathcal{P}_{2}:\  & \quad \max_{\substack{\mathbf{F}, \boldsymbol{\alpha}, \mathbf{c}}}
		\quad\quad t\label{prob2:formu_prob_sdp_a}                                                                                                                               \\
		\mbox{s.t.}        & \quad t\leq C_{g,n}+\alpha_{g,dn},\ n\in\mathcal{N}\label{prob2:formu_prob_sdp_b}                                                                   \\
		                   & \quad\sum_{n=1}^{N}C_{g,n}+\sum_{k=1}^{K}C_{l,k}^{sup}\leq \alpha_{g,cn},\ n\in\mathcal{N}\label{prob2:formu_prob_sdp_c}                            \\
		                   & \quad t\leq C_{l,k_m}^{sup}+C_{l,k_m}^{sub}+\alpha_{p,k_m},\ k_m\in\mathcal{K}_m,m\in\mathcal{M}\label{prob2:formu_prob_sdp_d}                      \\
		                   & \quad\sum_{n=1}^{N}C_{g,n}+\sum_{k=1}^{K}C_{l,k}^{sup}\leq \alpha_{l,k_m}^{sup},\ k_m\in\mathcal{K}_m,m\in\mathcal{M}\label{prob2:formu_prob_sdp_e} \\
		                   & \quad\sum_{k_m=1}^{K_m}C_{l,k_m}^{sub}\leq \alpha_{l,k_m}^{sub},\ k_m\in\mathcal{K}_m,m\in\mathcal{M}\label{prob2:formu_prob_sdp_f}                 \\
		                   & \quad \tr(\mathbf{F}_{g,c})+\tr(\mathbf{F}_{g,d})\leq P_g\label{prob2:formu_prob_sdp_g}                                                             \\
		                   & \quad \tr(\mathbf{F}_{l,cm})+\sum_{k_m=1}^{K_m}\tr(\mathbf{F}_{l,pk_m})\leq P_l,\ m\in\mathcal{M}\label{prob2:formu_prob_sdp_h}                     \\
		                   & \quad \mathbf{F}_{g,c}\succeq 0,\mathbf{F}_{g,d}\succeq 0,\label{prob2:formu_prob_sdp_i}                                                            \\
		                   & \quad \mathbf{F}_{l,cm}\succeq 0,\ m\in\mathcal{M}                                                                                                  \\
		                   & \quad \mathbf{F}_{l,pk_m}\succeq 0,\ k_m\in\mathcal{K}_{m},\ m\in\mathcal{M}\label{prob2:formu_prob_sdp_j}                                          \\
		                   & \quad \rank(\mathbf{F}_{g,c})=1,\rank(\mathbf{F}_{g,d})=1,\label{prob2:formu_prob_sdp_k}                                                            \\
		                   & \quad \rank(\mathbf{F}_{l,cm})=1, m\in\mathcal{M}                                                                                                   \\
		                   & \quad \rank(\mathbf{F}_{l,pk_m})=1,\ k_m\in\mathcal{K}_m,m\in\mathcal{M}\label{prob2:formu_prob_sdp_l}                                              \\
		                   & \quad \mathbf{c}\succeq\mathbf{0},\label{prob2:formu_prob_sdp_m}
	\end{align}
\end{subequations}
where $\boldsymbol{\alpha}_{g,c}\triangleq{\left[\alpha_{g,c1},\allowbreak\ldots,\allowbreak\alpha_{g,cN}\right]}^T$, $\boldsymbol{\alpha}_{g,p}\triangleq{\left[\alpha_{g,p1},\allowbreak\ldots,\allowbreak\alpha_{g,dN}\right]}^T$, $\boldsymbol{\alpha}_{l}^{sup}\triangleq\allowbreak{\left[\alpha_{l,1}^{sup},\allowbreak\ldots,\allowbreak\alpha_{l,K}^{sup}\right]}^T$, $\boldsymbol{\alpha}_{l}^{sub}\triangleq\allowbreak{\left[\alpha_{l,1}^{sub},\allowbreak\ldots,\allowbreak\alpha_{l,K}^{sub}\right]}^T$, $\boldsymbol{\alpha}_{p}\triangleq\allowbreak{\left[\alpha_{p,1},\allowbreak\ldots,\allowbreak\alpha_{p,K}\right]}^T$ represent the lower bound of corresponding achievable rates. (\ref{prob2:formu_prob_sdp_g}) and (\ref{prob2:formu_prob_sdp_h}) are the equivalent transmit power constraints.
\par Due to the imperfect \gls{csit} deployment and nonconvex nature of problem (\ref{prob2:formu_prob_sdp}), we denote
\begin{equation}
	\begin{aligned}
		\mathbf{H}_{g,n}= & \mathbb{E}\{\mathbf{h}_{g,n}\mathbf{h}_{g,n}^H\}                             \\
		=                 & \diag(\hat{\mathbf{h}}_{g,n})\mathbf{X}_{g,n}\diag(\hat{\mathbf{h}}_{g,n}^H)
	\end{aligned}
\end{equation}
as symmetric \gls{geo} channel matrix, where $\mathbf{X}_{g,n}=\mathbb{E}\{\boldsymbol{\Phi}_{eg,n}\boldsymbol{\Phi}_{eg,n}^H\}$ is the correlation matrix of $\boldsymbol{\Phi}_{eg,n}$. Similarly, $\mathbf{H}_{l2g,mn}, \mathbf{H}_{l,mk_m}$ and $\mathbf{H}_{g2l,k_m}$ are
\begin{subequations}
	\begin{align}
		 & \mathbf{H}_{l2g,mn}=\diag(\hat{\mathbf{h}}_{l2g,mn})\mathbf{X}_{l,mn}\diag(\hat{\mathbf{h}}_{l2g,mn}^H),     \\
		 & \mathbf{H}_{l,mk_m}=\diag(\hat{\mathbf{h}}_{l,mk_m})\mathbf{X}_{l,mk_m}\diag(\hat{\mathbf{h}}_{l,mk_m}^H),   \\
		 & \mathbf{H}_{g2l,k_m}=\diag(\hat{\mathbf{h}}_{g2l,k_m})\mathbf{X}_{g,k_m}\diag(\hat{\mathbf{h}}_{g2l,k_m}^H),
	\end{align}
\end{subequations}
where $\mathbf{X}_{l,mn}=\mathbb{E}\{\boldsymbol{\Phi}_{el,mn}\boldsymbol{\Phi}_{el,mn}^H\}$, $\mathbf{X}_{l,mk_m}=\mathbb{E}\{\boldsymbol{\Phi}_{el,mk_m}\boldsymbol{\Phi}_{el,mk_m}^H\}$ and $\mathbf{X}_{g,k_m}=\mathbb{E}\{\boldsymbol{\Phi}_{eg,k_m}\boldsymbol{\Phi}_{eg,k_m}^H\}$.
The symmetric residual channel matrixes are defined as $\tilde{\mathbf{H}}_{g,n}=\mathbb{E}\{\tilde{\mathbf{h}}_{g,n}\tilde{\mathbf{h}}^{H}_{g,n}\}$,
$\tilde{\mathbf{H}}_{g2l,k_m}=\mathbb{E}\{\tilde{\mathbf{h}}_{g2l,k_m}\tilde{\mathbf{h}}^{H}_{g2l,k_m}\}$,
$\tilde{\mathbf{H}}_{l,mk_m}=\mathbb{E}\{\tilde{\mathbf{h}}_{l,mk_m}\tilde{\mathbf{h}}^{H}_{l,mk_m}\}$.
Based on the symmetric channel matrixes and beamforming matrixes, the power segments of the received signal can be reformulated as
\begin{subequations}
	\begin{align}
		 & S_{g,jn}^\prime=      \tr\left(\mathbf{H}_{g,n}\mathbf{F}_{g,j}\right), j\in\{c,d\},                                                                             \\
		 & \tilde{S}_{g,jn}^\prime=      \tr(\tilde{\mathbf{H}}_{g,n}\mathbf{F}_{g,j}), j\in\{c,d\},                                                                        \\
		 & S_{g2l,ck_m}^\prime=  \tr\left(\mathbf{H}_{g2l,k_m}\mathbf{F}_{g,c}\right),\tilde{S}_{g2l,ck_m}^\prime=  \tr(\tilde{\mathbf{H}}_{g2l,k_m}\mathbf{F}_{g,c}),      \\
		 & S_{l,ck_m}^\prime=    \tr\left(\mathbf{H}_{l,mk_m}\mathbf{F}_{l,cm}\right),\tilde{S}_{l,ck_m}^\prime=    \tr(\tilde{\mathbf{H}}_{l,mk_m}\mathbf{F}_{l,cm}),      \\
		 & S_{l,pk_m}^\prime=    \tr\left(\mathbf{H}_{l,mk_m}\mathbf{F}_{l,pk_m}\right), \tilde{S}_{l,pk_m}^\prime=    \tr(\tilde{\mathbf{H}}_{l,mk_m}\mathbf{F}_{l,pk_m}), \\
		 & {I_{g,n}^{tot}}^\prime=     \sum_{m=1}^{M}\tr\left(\mathbf{H}_{l2g,mn}\mathbf{F}_{l,cm}\right)\notag                                                             \\
		 & \quad\quad\quad+\sum_{m=1}^{M}\sum_{i=1}^{K_m}\tr\left(\mathbf{H}_{l2g,mn}\mathbf{F}_{l,pi}\right),                                                              \\
		\begin{split}
			& {I_{l,k_m}^{tot}}^\prime=   \sum_{\substack{i=1,                                                                                                                                        \\i\neq k_m}}^{K_m}\tr\left(\mathbf{H}_{l,mk_m}\mathbf{F}_{l,pi}\right)                                                                                                                                           +\sum_{\substack{j=1,                                                          \\j\neq m}}^{M}\tr\left(\mathbf{H}_{l,jk_m}\mathbf{F}_{l,cj}\right) \\
			& \quad\quad\quad+\sum_{\substack{j=1,                                                                                                                                              \\j\neq m}}^{M}\sum_{i=1}^{K_j}\tr\left(\mathbf{H}_{l,jk_m}\mathbf{F}_{l,pi}\right)+\tr\left(\mathbf{H}_{g2l,k_m}\mathbf{F}_{g,d}\right).
		\end{split}
	\end{align}
\end{subequations}
The ergodic rate is approximated by the following method as in \cite{9165811}, take the general rate expression as an example,
\begin{equation}
	\begin{aligned}
		R=\mathbb{E}\{\log_2(1+\frac{c_1}{c_2})\}\approx\log_2(1+\frac{\mathbb{E}\{c_1\}}{\mathbb{E}\{c_2\}}),
	\end{aligned}
\end{equation}
where $c_1$ and $c_2$ are nonnegative random variable.
Therefore, the rate expressions can be redefined and approximated as
\begin{subequations}
	\begin{align}
		 & R_{g,cn}^\prime\approx\log_2\left(\frac{S_{g,cn}^\prime+S_{g,dn}^\prime+{I_{g,n}^{tot}}^\prime+\sigma_n^2}{\tilde{S}_{g,cn}^\prime+S_{g,dn}^\prime+{I_{g,n}^{tot}}^\prime+\sigma_n^2}\right)                                                                                               \\
		 & R_{g,dn}^\prime\approx\log_2\left(\frac{\tilde{S}_{g,cn}^\prime+S_{g,dn}^\prime+{I_{g,n}^{tot}}^\prime+\sigma_n^2}{{\tilde{S}_{g,cn}^\prime+\tilde{S}_{g,dn}^\prime+I_{g,n}^{tot}}^\prime+\sigma_n^2}\right)                                                                               \\
		 & {R_{l,k_m}^{sup}}^\prime\approx\log_2\left(\frac{S_{g2l,ck_m}^\prime+S_{l,ck_m}^\prime+S_{l,pk_m}^\prime+{I_{l,k_m}^{tot}}^\prime+\sigma_{k_m}^2}{\tilde{S}_{g2l,ck_m}^\prime+S_{l,ck_m}^\prime+S_{l,pk_m}^\prime+{I_{l,k_m}^{tot}}^\prime+\sigma_{k_m}^2}\right)                          \\
		 & {R_{l,k_m}^{sub}}^\prime\approx\log_2\left(\frac{\tilde{S}_{g2l,ck_m}^\prime+S_{l,ck_m}^\prime+S_{l,pk_m}^\prime+{I_{l,k_m}^{tot}}^\prime+\sigma_{k_m}^2}{\tilde{S}_{g2l,ck_m}^\prime+\tilde{S}_{l,ck_m}^\prime+{S}_{l,pk_m}^\prime+{I_{l,k_m}^{tot}}^\prime+\sigma_{k_m}^2}\right)        \\
		 & R_{p,k_m}^\prime\approx\log_2\left(\frac{\tilde{S}_{g2l,ck_m}^\prime+\tilde{S}_{l,pk_m}^\prime+S_{l,pk_m}^\prime+{I_{l,k_m}^{tot}}^\prime+\sigma_{k_m}^2}{\tilde{S}_{g2l,ck_m}^\prime+\tilde{S}_{l,ck_m}^\prime+\tilde{S}_{l,pk_m}^\prime+{I_{l,k_m}^{tot}}^\prime+\sigma_{k_m}^2}\right).
	\end{align}
\end{subequations}

In order to solve the non-convex parts in the rate expressions, we further introduce slack variables $\boldsymbol{\eta}=\allowbreak\left\{\boldsymbol{\eta}_{g,c},\ \allowbreak\boldsymbol{\eta}_{g,p},\ \allowbreak\boldsymbol{\eta}_{l}^{sup},\ \allowbreak\boldsymbol{\eta}_{l}^{sub},\ \allowbreak\boldsymbol{\eta}_{p}\right\}$
and $\boldsymbol{\xi}=\allowbreak\left\{\boldsymbol{\xi}_{g,c},\ \allowbreak\boldsymbol{\xi}_{g,p},\ \allowbreak\boldsymbol{\xi}_{l}^{sup},\ \allowbreak\boldsymbol{\xi}_{l}^{sub},\ \allowbreak\boldsymbol{\xi}_{p}\right\}$, where $\boldsymbol{\eta}_{g,c}\triangleq{\left[\eta_{g,c1},\allowbreak\ldots,\allowbreak\eta_{g,cN}\right]}^T$,
$\boldsymbol{\eta}_{g,p}\triangleq{\left[\eta_{g,p1},\allowbreak\ldots,\allowbreak\eta_{g,dN}\right]}^T$,
$\boldsymbol{\eta}_{l}^{sup}\triangleq{\left[\eta_{l,1}^{sup},\allowbreak\ldots,\allowbreak\eta_{l,K}^{sup}\right]}^T$, $\boldsymbol{\eta}_{l}^{sub}\triangleq{\left[\eta_{l,1}^{sub},\allowbreak\ldots,\allowbreak\eta_{l,K}^{sub}\right]}^T$, $\boldsymbol{\eta}_{p}\triangleq{\left[\eta_{p,1},\allowbreak\ldots,\allowbreak\eta_{p,K}\right]}^T$. 
$\boldsymbol{\xi}_{g,c}\triangleq{\left[\xi_{g,c1},\allowbreak\ldots,\allowbreak\xi_{g,cN}\right]}^T$,
$\boldsymbol{\xi}_{g,p}\triangleq{\left[\xi_{g,p1},\allowbreak\ldots,\allowbreak\xi_{g,dN}\right]}^T$,
$\boldsymbol{\xi}_{l}^{sup}\triangleq{\left[\xi_{l,1}^{sup},\allowbreak\ldots,\allowbreak\xi_{l,K}^{sup}\right]}^T$, $\boldsymbol{\xi}_{l}^{sub}\triangleq{\left[\xi_{l,1}^{sub},\allowbreak\ldots,\allowbreak\xi_{l,K}^{sub}\right]}^T$, $\boldsymbol{\xi}_{p}\triangleq{\left[\xi_{p,1},\allowbreak\ldots,\allowbreak\xi_{p,K}\right]}^T$. 
By taking \gls{gu}-$n$'s lower bound of designated rate as an example, $\alpha_{g,dn}\leq R_{g,dn}^\prime$, it can be transformed as
\begin{subequations}
	\begin{align}
		\alpha_{g,dn}\ln2 & \leq\ln(\frac{\tilde{S}_{g,cn}^\prime+S_{g,dn}^\prime+{I_{g,n}^{tot}}^\prime+\sigma_n^2}{{\tilde{S}_{g,cn}^\prime+\tilde{S}_{g,dn}^\prime+I_{g,n}^{tot}}^\prime+\sigma_n^2}) \\
		                  & =\ln(\tilde{S}_{g,cn}^\prime+S_{g,dn}^\prime+{I_{g,n}^{tot}}^\prime+\sigma_n^2)\notag                                                                                        \\
		                  & \quad-\ln({{\tilde{S}_{g,cn}^\prime+\tilde{S}_{g,dn}^\prime+I_{g,n}^{tot}}^\prime+\sigma_n^2})                                                                               \\
		                  & =\eta_{g,dn}-\xi_{g,dn},
	\end{align}
\end{subequations}
where $\eta_{g,dn}$ and $\xi_{g,dn}$ represent the lower bound and upper bound of corresponding terms, respectively.
\par Based on the auxiliary variables, $\mathcal{P}_2$ can be rewritten as $\mathcal{P}_3$.
\begin{subequations}\label{prob3:formu_prob_sdp}
	\begin{align}
		\hspace{-10mm}
		\mathcal{P}_{3}:\   \quad & \max_{\substack{\mathbf{F}, \boldsymbol{\alpha}, \mathbf{c}                                                                                                                 \\ \boldsymbol{\eta},\boldsymbol{\xi}}}
		\quad\quad t\label{prob3:formu_prob_sdp_a}                                                                                                                                                              \\
		\mbox{s.t.}         \quad & \alpha_{g,cn}\ln2\leq\eta_{g,cn}-\xi_{g,cn},\ n\in\mathcal{N}\label{prob3:formu_prob_sdp_b}                                                                                 \\
		\quad                     & S_{g,cn}^\prime+S_{g,dn}^\prime+{I_{g,n}^{tot}}^\prime+\sigma_n^2\geq e^{\eta_{g,cn}}\label{prob3:formu_prob_sdp_c}                                                         \\
		\quad                     & \tilde{S}_{g,cn}^\prime+S_{g,dn}^\prime+{I_{g,n}^{tot}}^\prime+\sigma_n^2\leq e^{\xi_{g,cn}}\label{prob3:formu_prob_sdp_d}                                                  \\
		\quad                     & \alpha_{g,dn}\ln2\leq\eta_{g,dn}-\xi_{g,dn},\ n\in\mathcal{N}\label{prob3:formu_prob_sdp_e}                                                                                 \\
		\quad                     & \tilde{S}_{g,cn}^\prime+S_{g,dn}^\prime+{I_{g,n}^{tot}}^\prime+\sigma_n^2\geq e^{\eta_{g,dn}}\label{prob3:formu_prob_sdp_f}                                                 \\
		\quad                     & \tilde{S}_{g,cn}^\prime+\tilde{S}_{g,dn}^\prime+{I_{g,n}^{tot}}^\prime+\sigma_n^2\leq e^{\xi_{g,dn}}\label{prob3:formu_prob_sdp_g}                                          \\
		\quad                     & \alpha_{l,k_m}^{sup}\ln2\leq\eta_{l,k_m}^{sup}-{\xi_{l,k_m}^{sup}},\ k_m\in\mathcal{K}_{m},\ m\in\mathcal{M}\label{prob3:formu_prob_sdp_h}                                  \\
		\quad                     & S_{g2l,ck_m}^\prime+S_{l,ck_m}^\prime+S_{l,pk_m}^\prime+{I_{l,k_m}^{tot}}^\prime+\sigma_n^2\geq e^{\eta_{l,k_m}^{sup}}\label{prob3:formu_prob_sdp_i}                        \\
		\quad                     & \tilde{S}_{g2l,ck_m}^\prime+S_{l,ck_m}^\prime+S_{l,pk_m}^\prime+{I_{l,k_m}^{tot}}^\prime+\sigma_{k_m}^2\leq e^{{\xi_{l,k_m}^{sup}}}\label{prob3:formu_prob_sdp_j}           \\
		\quad                     & \alpha_{l,k_m}^{sub}\ln2\leq{\eta_{l,k_m}^{sub}}-{\xi_{l,k_m}^{sub}},\ k_m\in\mathcal{K}_{m},\ m\in\mathcal{M}\label{prob3:formu_prob_sdp_k}                                \\
		\quad                     & \tilde{S}_{g2l,ck_m}^\prime+S_{l,ck_m}^\prime+S_{l,pk_m}^\prime+{I_{l,k_m}^{tot}}^\prime+\sigma_{k_m}^2\geq e^{{\eta_{l,k_m}^{sub}}}\label{prob3:formu_prob_sdp_l}          \\
		\quad                     & \tilde{S}_{g2l,ck_m}^\prime+\tilde{S}_{l,ck_m}^\prime+{S}_{l,pk_m}^\prime+{I_{l,k_m}^{tot}}^\prime+\sigma_{k_m}^2\leq e^{{\xi_{l,k_m}^{sub}}}\label{prob3:formu_prob_sdp_m} \\
		\quad                     & \alpha_{p,k_m}\ln2\leq\eta_{p,k_m}-\xi_{p,k_m},\ k_m\in\mathcal{K}_{m},\ m\in\mathcal{M}\label{prob3:formu_prob_sdp_n}                                                      \\
		\quad                     & \tilde{S}_{g2l,ck_m}^\prime+\tilde{S}_{l,ck_m}^\prime+S_{l,pk_m}^\prime+{I_{l,k_m}^{tot}}^\prime+\sigma_{k_m}^2\geq e^{\eta_{p,k_m}}\label{prob3:formu_prob_sdp_o}          \\
		\quad                     & \tilde{S}_{g2l,ck_m}^\prime+\tilde{S}_{l,ck_m}^\prime+\tilde{S}_{l,pk_m}^\prime+{I_{l,k_m}^{tot}}^\prime+\sigma_{k_m}^2\leq e^{\xi_{p,k_m}}\label{prob3:formu_prob_sdp_p}   \\
		\quad                     & (\text{\ref{prob2:formu_prob_sdp_b}})-(\text{\ref{prob2:formu_prob_sdp_m}}).\notag
	\end{align}
\end{subequations}
\begin{figure*}[!t]
	\begin{equation}\label{equ:object_func3_approx}
		\begin{aligned}
			f_P & =  \beta\bigg\{\left[\tr(\mathbf{F}_{g,c})-{\left(\mathbf{v}_{\max g,c}^{[i]}\right)}^H\mathbf{F}_{g,c}{\mathbf{v}_{\max g,c}^{[i]}}\right]   + \left[\tr(\mathbf{F}_{g,d})-{\left(\mathbf{v}_{\max g,p}^{[i]}\right)}^H\mathbf{F}_{g,d}{\mathbf{v}_{\max g,p}^{[i]}}\right]                                                                       \\
			    & \quad + \sum_{m=1}^{M}\left[\tr(\mathbf{F}_{l,cm})-{\left(\mathbf{v}_{\max l,cm}^{[i]}\right)}^H\mathbf{F}_{l,cm}{\mathbf{v}_{\max l,cm}^{[i]}}\right]+ \sum_{m=1}^{M}\sum_{k_m=1}^{K_m}\bigg[\tr\left(\mathbf{F}_{l,pk_m}\right)-{\left(\mathbf{v}_{\max l,pk_m}^{[i]}\right)}^H\mathbf{F}_{l,pk_m}{\mathbf{v}_{\max l,pk_m}^{[i]}}\bigg]\bigg\}.
		\end{aligned}
	\end{equation}
	\hrulefill
\end{figure*}
Note that (\ref{prob3:formu_prob_sdp_d}), (\ref{prob3:formu_prob_sdp_g}), (\ref{prob3:formu_prob_sdp_j}), (\ref{prob3:formu_prob_sdp_m}) and (\ref{prob3:formu_prob_sdp_p}) are nonconvex with convex \gls{rhs} which can be approximated by the first-order Taylor series. The \gls{rhs} parts of the above constraints are approximated at $\Big\{\xi_{g,cn}^{[i]},\allowbreak\xi_{g,dn}^{[i]},\allowbreak{\xi_{l,k_m}^{sup}}^{[i]},\allowbreak{\xi_{l,k_m}^{sub}}^{[i]},\allowbreak\xi_{p,k_m}^{[i]}|n\in\mathcal{N},\allowbreak\ k_m\in\mathcal{K}_m,\allowbreak\ m\in\mathcal{M}\Big\}$ at iteration $i$ as (\ref{equ:xi_approx2})

\begin{equation}\label{equ:xi_approx2}
	e^{j}\leq e^{j^{[i]}}\left[j-j^{[i]}+1\right],
\end{equation}
where $j\in\left\{\xi_{g,cn},\xi_{g,dn},{\xi_{l,k_m}^{sup}},{\xi_{l,k_m}^{sub}},\xi_{p,k_m}\right\}$. Hence, the constraints (\ref{prob3:formu_prob_sdp_d}), (\ref{prob3:formu_prob_sdp_g}), (\ref{prob3:formu_prob_sdp_j}), (\ref{prob3:formu_prob_sdp_m}) and (\ref{prob3:formu_prob_sdp_p}) can be rewritten as

\begin{subequations}\label{equ:xi_approx}
	\begin{align}
		 & \tilde{S}_{g,cn}^\prime+S_{g,dn}^\prime+{I_{g,n}^{tot}}^\prime+\sigma_n^2 \notag                                               \\
		 & \quad\leq e^{\xi_{g,cn}^{[i]}}\left[\xi_{g,cn}-\xi_{g,cn}^{[i]}+1\right]\label{equ:xi_approx_a}                                \\
		 & \tilde{S}_{g,cn}^\prime+\tilde{S}_{g,dn}^\prime+{I_{g,n}^{tot}}^\prime+\sigma_n^2\notag                                        \\
		 & \quad\leq e^{\xi_{g,dn}^{[i]}}\left[\xi_{g,dn}-\xi_{g,dn}^{[i]}+1\right]\label{equ:xi_approx_b}                                \\
		 & \tilde{S}_{g2l,ck_m}^\prime+S_{l,ck_m}^\prime+S_{l,pk_m}^\prime+{I_{l,k_m}^{tot}}^\prime+\sigma_{k_m}^2\notag                  \\
		 & \quad\leq e^{{\xi_{l,k_m}^{sup}}^{[i]}}\left[{\xi_{l,k_m}^{sup}}-{\xi_{l,k_m}^{sup}}^{[i]}+1\right]\label{equ:xi_approx_c}     \\
		 & \tilde{S}_{g2l,ck_m}^\prime+\tilde{S}_{l,ck_m}^\prime+{S}_{l,pk_m}^\prime+{I_{l,k_m}^{tot}}^\prime+\sigma_{k_m}^2\notag        \\
		 & \quad\leq e^{{\xi_{l,k_m}^{sub}}^{[i]}}\left[{\xi_{l,k_m}^{sub}}-{\xi_{l,k_m}^{sub}}^{[i]}+1\right]                            \\
		 & \tilde{S}_{g2l,ck_m}^\prime+\tilde{S}_{l,ck_m}^\prime+\tilde{S}_{l,pk_m}^\prime+{I_{l,k_m}^{tot}}^\prime+\sigma_{k_m}^2 \notag \\
		 & \quad\leq e^{\xi_{p,k_m}^{[i]}}\left[\xi_{p,k_m}-\xi_{p,k_m}^{[i]}+1\right]\label{equ:xi_approx_e}.
	\end{align}
\end{subequations}

\par Rank-one implies that there is only one nonzero eigenvalue, hence, the nonconvex constraints (\ref{prob2:formu_prob_sdp_k}) and (\ref{prob2:formu_prob_sdp_l}) can be reexpressed as
\begin{subequations}\label{equ:rank_one_rewritre}
	\begin{align}
		 & \tr(\mathbf{F}_{g,c})-\lambda_{\max}\left(\mathbf{F}_{g,c}\right)=0,\ \tr(\mathbf{F}_{g,d})-\lambda_{\max}\left(\mathbf{F}_{g,d}\right)=0, \\
		 & \tr(\mathbf{F}_{l,cm})-\lambda_{\max}\left(\mathbf{F}_{l,cm}\right)=0,\ m\in\mathcal{M},                                                   \\
		 & \tr\left(\mathbf{F}_{l,pk_m}\right)-\lambda_{\max}\left(\mathbf{F}_{l,pk_m}\right)=0,\ k_m\in\mathcal{K}_m,\ m\in\mathcal{M}.
	\end{align}
\end{subequations}
In order to include these constraints into the objective function (\ref{prob3:formu_prob_sdp_a}), we construct a penalty function and obtain
\begin{equation}\label{equ:object_func1}
	\begin{aligned}
		\max_{\substack{\mathbf{F}, \boldsymbol{\alpha}, \mathbf{c}                                                                                          \\ \boldsymbol{\eta},\boldsymbol{\xi}}}
		\quad   t & -\beta\Big\{  \left[\tr(\mathbf{F}_{g,c})-\lambda_{\max}\left(\mathbf{F}_{g,c}\right)\right]                                             \\
		          & + \left[\tr(\mathbf{F}_{g,d})-\lambda_{\max}\left(\mathbf{F}_{g,d}\right)\right]                                                         \\
		          & + \sum_{m=1}^{M}\left[\tr(\mathbf{F}_{l,cm})-\lambda_{\max}\left(\mathbf{F}_{l,cm}\right)\right]                                         \\
		          & + \sum_{m=1}^{M}\sum_{k_m=1}^{K_m}\left[\tr\left(\mathbf{F}_{l,pk_m}\right)-\lambda_{\max}\left(\mathbf{F}_{l,pk_m}\right)\right]\Big\}.
	\end{aligned}
\end{equation}
To ensure that the penalty function is as small as possible, we choose the appropriate penalty factor $\beta$. Due to the existence of the penalty function, (\ref{equ:object_func1}) is not concave. We utilize an iterative approach to solve this problem \cite{9165811}. Using $\tr(\mathbf{F}_{g,c})-\lambda_{\max}\left(\mathbf{F}_{g,c}\right)$ as an example, the following inequality can be derived
\begin{equation}\label{equ:object_func2_approx}
	\begin{aligned}
		0 & \leq\tr(\mathbf{F}_{g,c})-\lambda_{\max}\left(\mathbf{F}_{g,c}\right)                                                \\
		  & \leq\tr(\mathbf{F}_{g,c})-{\left(\mathbf{v}_{\max g,c}^{[i]}\right)}^H\mathbf{F}_{g,c}{\mathbf{v}_{\max g,c}^{[i]}},
	\end{aligned}
\end{equation}
where $\mathbf{v}_{\max g,c}$ is a normalized eigenvector in association with the largest eigenvalue $\lambda_{\max}\left(\mathbf{F}_{g,c}\right)$. Likewise, we define $\mathbf{v}_{\max g,p}$ for $\lambda_{\max}\left(\mathbf{F}_{g,d}\right)$ and $\mathbf{v}_{\max l,cm}$ for $\lambda_{\max}\left(\mathbf{F}_{l,cm}\right)$ as well as $\mathbf{v}_{\max l,pk_m}$ for $\lambda_{\max}\left(\mathbf{F}_{l,pk_m}\right)$'s associated eigenvectors. The iterative penalty function is represented by $f_P$ in (\ref{equ:object_func3_approx}).

Therefore, we solve the following subproblem:

\begin{subequations}\label{prob4:formu_prob_sdp}
	\begin{align}
		\mathcal{P}_{4}:\   \quad  \max_{\substack{\substack{\mathbf{F}, \boldsymbol{\alpha}, \mathbf{c}                                                                                                                              \\ \boldsymbol{\eta},\boldsymbol{\xi}}}}
		\quad\quad             & t-f_P\label{prob4:formu_prob_sdp_a}                                                                                                                                                                  \\
		\mbox{s.t.} \quad\quad & (\text{\ref{prob2:formu_prob_sdp_b}})-(\text{\ref{prob2:formu_prob_sdp_j}}),(\text{\ref{prob2:formu_prob_sdp_m}}),(\text{\ref{prob3:formu_prob_sdp_b}}),\notag                                       \\
		                       & (\text{\ref{prob3:formu_prob_sdp_c}}),(\text{\ref{prob3:formu_prob_sdp_e}}),(\text{\ref{prob3:formu_prob_sdp_f}}),(\text{\ref{prob3:formu_prob_sdp_h}}),(\text{\ref{prob3:formu_prob_sdp_i}}),\notag \\
		                       & (\text{\ref{prob3:formu_prob_sdp_k}}),(\text{\ref{prob3:formu_prob_sdp_l}}),(\text{\ref{prob3:formu_prob_sdp_n}}),(\text{\ref{prob3:formu_prob_sdp_o}}),(\ref{equ:xi_approx}).\notag
	\end{align}
\end{subequations}

Problem $\mathcal{P}_{1}$ has been transformed into convex problem and can be solved effectively using the CVX toolbox \cite{grant2008cvx}. While solving $\mathcal{P}_{4}$, the results $\left\{{\mathbf{F}}^{[i]},\boldsymbol{\xi}^{[i]}\right\}$ from the $i$-th iteration are considered as constants. The existence of lower bounds, i.e., (\ref{equ:object_func2_approx}), ensures that the objective function converges. In other words, the rank-one constraints can be satisfied \cite{9165811}. If the optimized symmetric matrices, $\big\{\mathbf{F}_{g,c},\allowbreak\mathbf{F}_{g,p},\allowbreak\mathbf{F}_{l,cm},\allowbreak\mathbf{F}_{l,pk}|\allowbreak k\in\mathcal{K},\allowbreak m\in\mathcal{M}\big\}$, are of rank one, then the optimal precoders of \gls{geo} and \gls{leo} satellites can be obtained by using eigenvalue decomposition. Alternatively, randomization can be adopted to extract approximate solutions from the optimized symmetric matrices, however, this leads to higher complexity.
\par The steps of multilayer \gls{drsma} joint \gls{geo}-\gls{leo} precoders optimization scheme is summarized in Algorithm \ref{Algor:DRS_SDP}.
The proposed robust beamforming design for \gls{drsma} is based on statistical \gls{csi}, which means that the optimization problem can be optimized offline, and the optimized precoders can be implemented to all channel uses. The proposed algorithm provides a theoretical upper bound for the system, since it is based on the assumptions that physical channel parameters are invariant over the time intervals of interest and time/frequency synchronization can be properly performed at each UT.
The precoders are initialized by using \gls{mrt} and \gls{svd} since they have been demonstrated to provide good overall performance over a variety of channel realizations \cite{Mao2018,7555358}.
\begin{algorithm}[t!]
	\caption{\gls{sdp}-based Optimization Scheme}\label{Algor:DRS_SDP}
	\textbf{Initialize}: $i\leftarrow0$, $t^{[i]}\leftarrow0$, ${\mathbf{F}}^{[i]}$, $\boldsymbol{\xi}^{[i]}$\;
	\Repeat{$|\{t-f_P\}^{[i]}-\{t-f_P\}^{[i-1]}|<\tau$}{
	$i\leftarrow i+1$\;
	{Solve problem (\ref{prob4:formu_prob_sdp}) using ${\mathbf{F}}^{[i-1]}$, $\boldsymbol{\xi}^{[i-1]}$ and denote the optimal value of the objective function as $\{t-f_P\}^{[i]}$ and the optimal solutions as ${\mathbf{F}}^{[i]}$, $\boldsymbol{\xi}^{[i]}$}\;
	}
	\For{$j\in\big\{\{g,c\},\{g,p\},\{l,cm\},\{l,pk\}|k\in\mathcal{K},\allowbreak m\in\mathcal{M}$\big\}}{
		\eIf{$\rank(\mathbf{F}_j)=1$}{Use eigenvalue decomposition to $\mathbf{F}_j$ and obtain the corresponding precoder}{Use randomization to extract an approximate solution}}
\end{algorithm}
The tolerance of the algorithm is denoted by $\tau=10^{-5}$.
\par The convergence of Algorithm \ref{Algor:DRS_SDP} is guaranteed since the solutions to the problem (\ref{prob4:formu_prob_sdp}) at iteration $i-1$ is a feasible solution to the problem at iteration $i$. Therefore, the objective function $t-f_P$ rises monotonically and it is constrained above by the transmit power. At each iteration, the solution meets the Karush-Kuhn-Tucker (KKT) optimality criteria, which are the same as those of (\ref{prob:formu_prob_orig}) at convergence \cite{8642812}. The total number of iterations required for the convergence is approximated as $\mathcal{O}(\log(\tau^{-1}))$.
At each iteration of the proposed \gls{sdp}-based algorithm, the convex subproblem $\mathcal{P}_{4}$ is solved. The computational complexity of each iteration is mostly determined by the number of optimization variables, the number and size of \gls{lmi} constraints, and the size of \gls{soc} constraints \cite{9180053,ben2001lectures}. The problem $\mathcal{P}_{4}$ has $2N_{tg}^2+(K+M)N_{tl}^2+2K+N$ design variables, $6N+9K+1$ slack variables, and $8N+12K+M+1$ \gls{lmi} constraints of size 1. Therefore, the worst-case computational complexity is given by

\begin{equation}
	\begin{split}
		\mathcal{O} & \bigg(\log(\tau^{-1}){(8N+12K+M+1)}^{1/2}\cdot z\Big[z^2\\
		&+2N_{tg}^2(N_{tg}+z)+(K+M)N_{tl}^2(N_{tl}+z)\\
		&+(2K+N)(1+z)+(8N+12K+M+1)(1+z)\Big]\bigg),
	\end{split}
\end{equation}
where $z=\mathcal{O}\left(2N_{tg}^2+(K+M)N_{tl}^2+2K+7N+9K\right)$.

\section{Simulation Results}\label{sec:Simu_results}
The performance of the proposed \gls{drsma} scheme is evaluated in this section. We assume all \gls{geo} and \gls{leo} satellites operate in the Ka-band \cite{38811,38821,9257433}. The major simulation parameters are listed in Table \ref{tab:simu_para}. Specifically, we assume the \gls{geo} satellite is equipped with $N_{tg} = 4$ antennas, and $N = 6$ multicasting \gls{gus} locate uniformly in coverage area. Each \gls{leo} satellite is deployed with $N_{tl}=3$ antennas, and \gls{lus} are uniformly distributed within \gls{leo} coverage.
Because the noise power is normalized by $\kappa T_{sys} B_w$ in (\ref{equ:GEO_channel}) and (\ref{equ:LEO_channel_sa_gain}), we denote unit noise variance, i.e., $\sigma_r^2=\sigma_n^2=\sigma_{k_m}^2=1$, $\forall n\in\mathcal{N}$, $\forall k_m\in\mathcal{K}$, $m\in\mathcal{M}$. We redefine $P_l$ as \gls{snr} of \gls{leo} satellite and $P_g$ as \gls{snr} of \gls{geo} satellite.
We first investigate the scenario with perfect \gls{csi} in the network, then we move to the deployment with imperfect \gls{csit} and imperfect \gls{csir}. All \gls{mmf} simulation results are obtained by averaging $100$ channel realizations.
\par We compare the \gls{drsma} with three baseline multiple access schemes, namely, ``M-RSMA'', ``M-SDMA'' and  ``M-NOMA''. ``M-RSMA'' means multicasting is adopted at the \gls{geo} satellite and \gls{leo} satellites adopt 1-layer \gls{rsma}, while ``M-SDMA'' and ``M-NOMA'' denote multicasting is adopted at the \gls{geo} and \gls{leo} satellites implement \gls{sdma} and \gls{noma}, respectively.

\begin{table}[!t]
	\caption{Simulation Parameters}
	\begin{tabular}{IlIcIcI}
		\bottomrule[1.05pt]
		Parameter                  & \gls{geo}                                                            & \gls{leo}  \\ \toprule[1.05pt]
		\bottomrule[1.05pt]
		Carrier frequency          & \multicolumn{2}{cI}{20 GHz}                                                       \\ \hline
		Bandwidth                  & \multicolumn{2}{cI}{500 MHz}                                                      \\ \hline
		User terminal antenna gain & \multicolumn{2}{cI}{39.7 dBi}                                                     \\ \hline
		Boltzmann constan          & \multicolumn{2}{cI}{$1.38\times10^{-23}\text{J}\cdot \text{K}^{-1}$}              \\ \hline
		Noise temperature          & \multicolumn{2}{cI}{$290$ K}                                                      \\ \hline
		Satellite height           & 35786 km                                                             & 600 km     \\ \hline
		Antenna gain               & 58.5 dBi                                                             & 30.5 dBi   \\ \hline
		3 dB bandwidth             & 0.4412 deg                                                           & 4.4127 deg \\ \hline
		Rician factor              &                                                                      & 10 dB      \\ \toprule[1.05pt]
	\end{tabular}
	\label{tab:simu_para}
\end{table}

\begin{figure}[!t]
	\centering
	\includegraphics[scale=0.6]{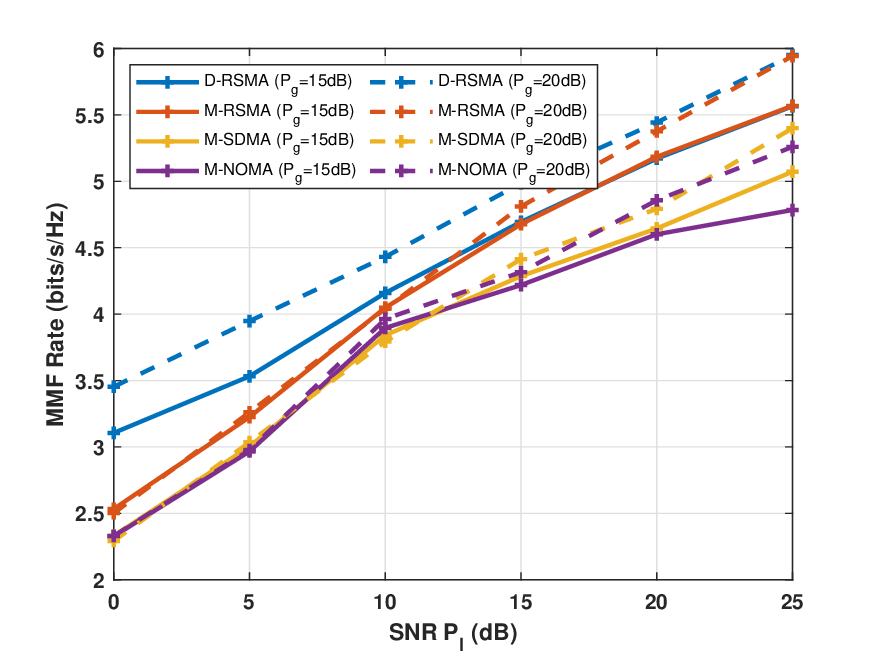}
	\caption{MMF rate versus \gls{leo} SNR $P_l$ for different \gls{geo} SNR $P_g$ and network loads. $N_{tg}=4$, $N_{tl}=3$, $M=2$, $N=6$, $K=4$.}
	\label{fig:rate_comp_1}
\end{figure}
\begin{figure}[!t]
	\centering
	\includegraphics[scale=0.6]{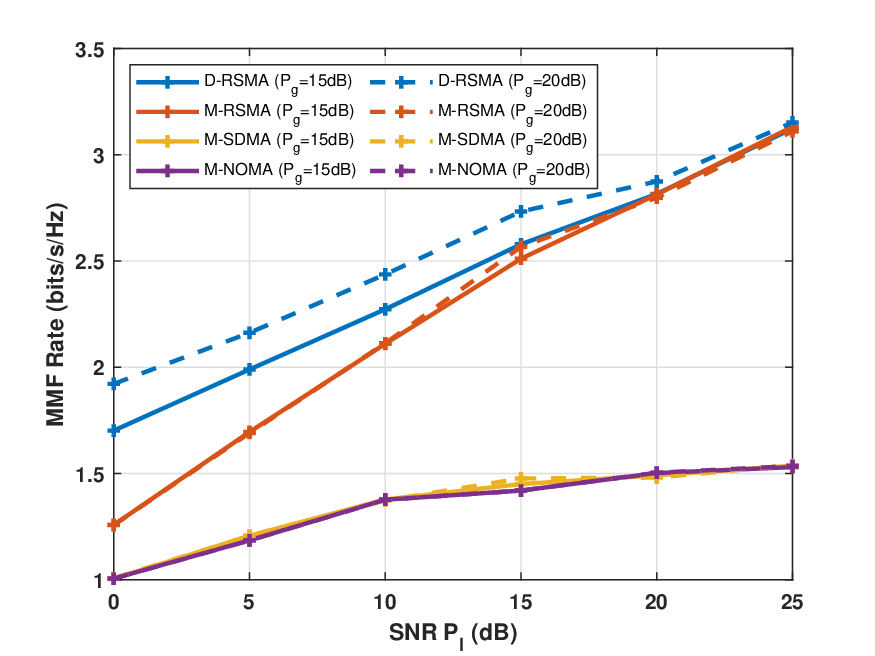}
	\caption{MMF rate versus \gls{leo} SNR $P_l$ for different \gls{geo} SNR $P_g$ and network loads. $N_{tg}=4$, $N_{tl}=3$, $M=2$, $N=6$, $K=8$.}
	\label{fig:rate_comp_2}
\end{figure}

Fig. \ref{fig:rate_comp_1} and \ref{fig:rate_comp_2} compare the \gls{mmf} performance of \gls{drsma} and baseline schemes under different system setups. Fig. \ref{fig:rate_comp_1} shows the \gls{mmf} performance under $2$ \gls{leo} satellites and $4$ \gls{lus} deployments, each \gls{leo} satellite serves $2$ \gls{lus}. Fig. \ref{fig:rate_comp_2} illustrates the \gls{mmf} rate for overloaded scenario, i.e., $8$ \gls{lus} in total, and $4$ \gls{lus} are served by one \gls{leo} satellite.
From both figures, it can be seen that 
\gls{drsma} achieves a clear \gls{mmf} rate gain over all the other schemes. 
As $P_l$ increases, the gap between \gls{drsma} and M-RSMA schemes gradually decreases and the two schemes overlap when $P_l$ is large enough. The reasons are as follows. The joint \gls{geo}-\gls{leo} beamforming optimization method is designed to attain the optimized \gls{mmf} rates. When the power budget of \gls{leo} is relatively small, the system performance is restricted by the \gls{lus} since the \gls{sinr}s of \gls{lus} are much lower than those of \gls{gus}. However, in the \gls{drsma} scheme, \gls{lus} messages are divided and parts of \gls{lus}' data are encoded into common stream $s_{g,c}$ transmitted from the \gls{geo} satellite (streams are transmitted from different satellites, thereby \gls{rsma} is distributedly implemented). This enables \gls{drsma} to utilize \gls{geo} power to transmit parts of \gls{lus} data, balancing the network load and managing the interference between two sub-networks.
Due to the fixed \gls{geo} satellite transmit power budget $P_g$, the \gls{mmf} rates of both \gls{drsma} and M-RSMA saturate when $P_l$ is sufficiently large. The benefit of the \gls{drsma} over the other strategies decreases as $P_l$ increases. M-RSMA outperforms M-SDMA because we consider the \gls{los} channels and the \gls{lus} channels are closely aligned with each other. 
Compared to Fig. \ref{fig:rate_comp_1}, the attained \gls{mmf} rate in Fig. \ref{fig:rate_comp_2} decreases, while the benefit of distributed \gls{rsma} is enhanced further.
The relative gain of ``\gls{drsma} ($P_g=15$ dB)'' over ``M-RSMA ($P_g=15$ dB)'' increases from $2.74\%$ in Fig. \ref{fig:rate_comp_1} to $7.71\%$ in Fig. \ref{fig:rate_comp_2} when $P_l=10$ dB.
With more served \gls{lus}, the simultaneous inter-system and intra-system interference are severer. M-RSMA cannot manage the interference between sub-networks. However, by transmitting $s_{g,c}$ from \gls{geo}, the interference can be coordinated and suppressed with the implementation of \gls{drsma}.



\par The influence of the \gls{geo} power budget is also investigated. From Fig. \ref{fig:rate_comp_1}, the larger $P_g$ leads to better \gls{mmf} rate performance. Besides, \gls{drsma} benefits more from the increase in power budget since \gls{geo} can allocate more power to transmit the common message. 
Hence the overall max-min rate can be improved and the saturated rate when $P_l$ is sufficiently large increases compared to that in the scenario with $P_g=15$ dB.
When \gls{leo} satellites work in the overloaded deployment (Fig. \ref{fig:rate_comp_2}), only \gls{drsma} benefits from the increase of $P_g$ since it can make use of the power from both sub-networks and manage the severer interference between \gls{geo} and \gls{leo} sub-networks. 
We can conclude that \gls{drsma} can well manage the interference and
achieves \gls{mmf} rate gain compared to M-RSMA and M-SDMA regardless of network loads, which guarantees user fairness. 
Alternatively, \gls{drsma} can utilize less power to achieve the same \gls{mmf} performance.


\begin{figure}[t]
	\centering
	\includegraphics[scale=0.6]{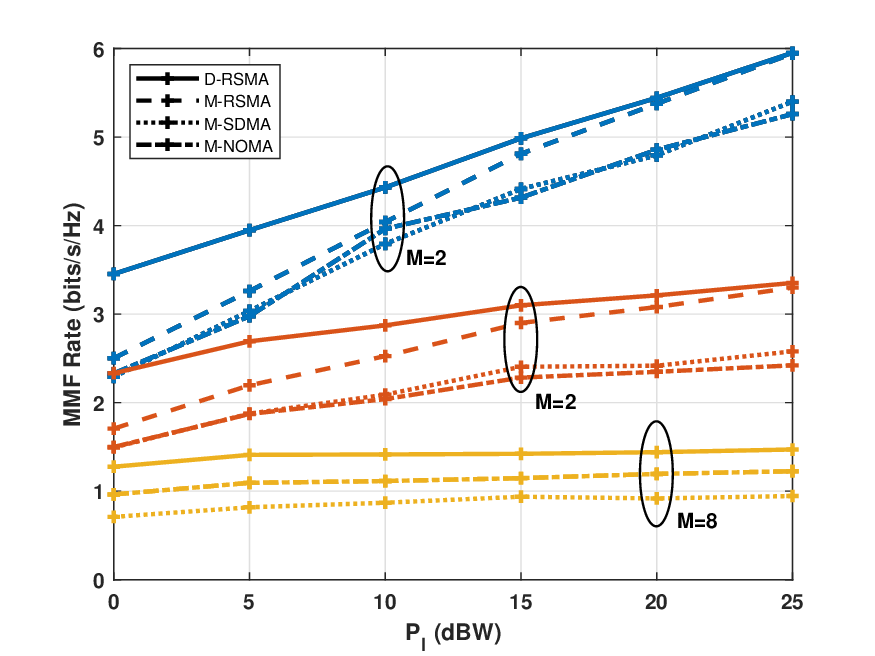}
	\caption{\gls{mmf} rate versus \gls{leo} SNR $P_l$ with varied number of \gls{leo} satellites. $N_{tg}=4$, $N_{tl}=3$, $N=6$, $P_g=20$ dB.}
	\label{fig:NumLeo_comp}
\end{figure}

\par The influence of the number of \gls{leo} satellites is illustrated in Fig. \ref{fig:NumLeo_comp}. We illustrate the performance of proposed scheme in dense \gls{leo} networks with $M=2$, $M=4$ and $M=8$ \gls{leo} satellites scenarios.
$2$ \gls{lus} are served by each \gls{leo} satellite.
By increasing the number of \gls{leo} satellites and \gls{lus}, the \gls{mmf} performances of all schemes degrade, while \gls{drsma} still outperforms the other baseline approaches in all deployments.
The relative gain of \gls{drsma} over M-RSMA increases from $9.68\%$ in $2$ \gls{leo} satellites deployment to $13.74\%$ in $4$ \gls{leo} satellites and to $37.41\%$ in $8$ \gls{leo} satellites deployment when $P_l=10$ dB.
Because, as the increase of \gls{leo} satellites and \gls{lus}, each UT suffers from severer inter-system and intra-system interference, without the transmission of $s_{g,c}$, M-RSMA cannot manage the interference coming from \gls{gus} and \gls{lus} that are not served by the same \gls{leo}. In comparison, \gls{drsma} is more robust to the number of co-existence \gls{leo} satellites and network load thanks to its ability to partially decode the interference and partially treat interference as noise.

\begin{figure}[t]
	\centering
	\includegraphics[scale=0.6]{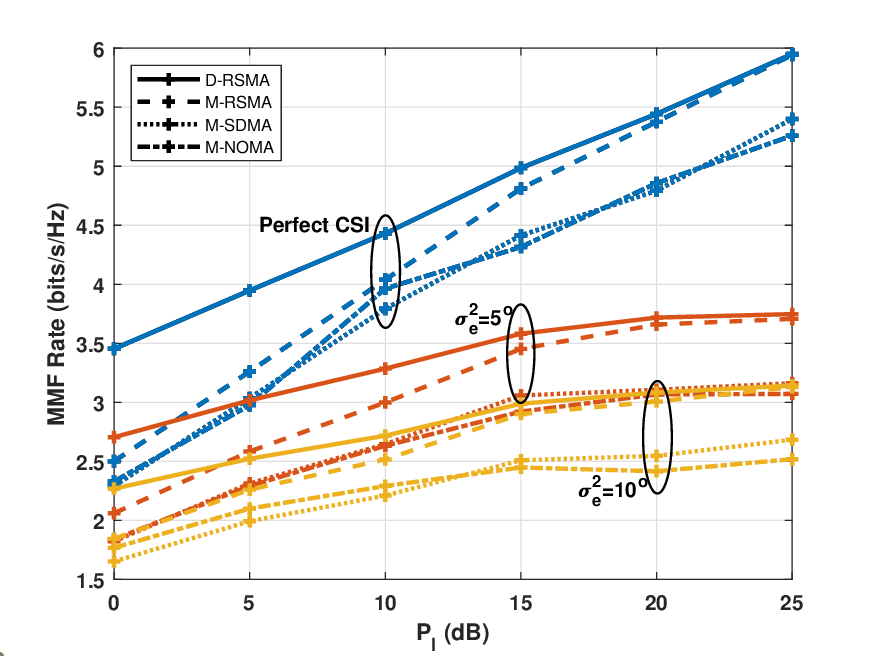}
	\caption{\gls{mmf} rate versus \gls{leo} SNR $P_l$ with varied satellite phase uncertainties. $M=2$, $N_{tg}=4$, $N_{tl}=3$, $N=6$, $K=4$, $P_g=20$ dB.}
	\label{fig:CSITR_comp}
\end{figure}
\par Furthermore, we take phase uncertainty into account. We assume $\sigma_{e,g}^2=\sigma_{e,l}^2=\sigma_{e}^2$. The ergodic \gls{mmf} performance of the proposed method and baselines under imperfect \gls{csit} and imperfect \gls{csir} is investigated in Fig. \ref{fig:CSITR_comp}. The \gls{mmf} rates in all approaches decrease with a rise in phase uncertainty variation, while \gls{drsma} outperforms baselines in all scenarios. When $P_l=20$ dB, from perfect CSI (i.e., $\sigma_e^2=0^{\circ}$) to the deployment with $\sigma_{e}^2=5^{\circ}$ and $\sigma_{e}^2={10}^{\circ}$, the corresponding ergodic \gls{mmf} rate decreases by $30.69\%$ and $42.35\%$ respectively for \gls{drsma}, and decreases by $32.93\%$ and $45.06\%$ respectively for M-SDMA. For comparison, the ergodic \gls{mmf} rate of M-SDMA decreases by $36.19\%$ and $47.86\%$ from perfect CSI to the deployment with $\sigma_{e}^2=5^{\circ}$ and $\sigma_{e}^2={10}^{\circ}$, respectively.
\gls{drsma} and M-RSMA have a more flexible design that allows them to partially decode interference and treat it as noise. Therefore \gls{drsma} is more robust to channel phase uncertainty.

\section{Conclusion}\label{sec:Conclu}
To conclude, we have investigated the co-existence of \gls{geo}-\gls{leo} multilayer network with statistical \gls{csit} and \gls{csir}, where the \gls{rsma} is distributedly implemented (\gls{drsma}) to mitigate the interference and improve user fairness. To this end, we have formulated a \gls{mmf} problem that jointly optimizes the precoders of \gls{geo} and \gls{leo} satellites and the message split of each user subject to the power constraints at the satellites. We have proposed a robust \gls{sdp}-based iterative optimization algorithm to solve this problem. We have further showed show through numerical results that the proposed \gls{drsma} can well manage the interference and boost the \gls{mmf} performance of the system. The influence of the \gls{geo} power budget, the number of \gls{lus}, the number of \gls{leo} satellites and \gls{csi} uncertainty have been studied. We have found that \gls{drsma} is robust to the channel phase uncertainty. Besides, as the increase of \gls{geo} transmit power, or the number of serving \gls{lus}, the rate improvement of \gls{drsma} over other \gls{rsma}, \gls{sdma} and \gls{noma} baselines become more significant.
Thanks to the transmission of \gls{geo} common stream, \gls{drsma} is more capable of managing inter-system and intra-system interference. These results lead us to conclude that \gls{drsma} is superior to existing transmission schemes and has a great potential to improve the system performance of future communication networks.


\bibliographystyle{IEEEtran}
\bibliography{All_ref.bib}

\begin{thebibliography}{10}
\providecommand{\url}[1]{#1}
\csname url@samestyle\endcsname
\providecommand{\newblock}{\relax}
\providecommand{\bibinfo}[2]{#2}
\providecommand{\BIBentrySTDinterwordspacing}{\spaceskip=0pt\relax}
\providecommand{\BIBentryALTinterwordstretchfactor}{4}
\providecommand{\BIBentryALTinterwordspacing}{\spaceskip=\fontdimen2\font plus
\BIBentryALTinterwordstretchfactor\fontdimen3\font minus \fontdimen4\font\relax}
\providecommand{\BIBforeignlanguage}[2]{{%
\expandafter\ifx\csname l@#1\endcsname\relax
\typeout{** WARNING: IEEEtran.bst: No hyphenation pattern has been}%
\typeout{** loaded for the language `#1'. Using the pattern for}%
\typeout{** the default language instead.}%
\else
\language=\csname l@#1\endcsname
\fi
#2}}
\providecommand{\BIBdecl}{\relax}
\BIBdecl

\bibitem{22822}
{3GPP TR 22.822 V16.0.0}, ``{3rd Generation Partnership Project; Technical Specification Group Services and System Aspects; Study on using Satellite Access in 5G; Stage 1 (Release 16)},'' 3GPP, techreport, Jun. 2018.

\bibitem{38821}
{3GPP TR 38.821 V16.1.0}, ``3rd {Generation Partnership Project}; {Technical Specification Group Radio Access Network; Solutions for NR to support non-terrestrial networks (NTN) (Release 16)},'' 3GPP, techreport, May 2021.

\bibitem{38811}
{3GPP TR 38.811 V15.4.0}, ``3rd {Generation Partnership Project}; {Technical Specification Group Radio Access Network; Study on New Radio (NR) to support non-terrestrial networks (Release 15)},'' 3GPP, techreport, Sep. 2020.

\bibitem{9815078}
Y.~Liu, Y.~Wang, J.~Wang, L.~You, W.~Wang, and X.~Gao, ``Robust downlink precoding for {LEO} satellite systems with per-antenna power constraints,'' \emph{IEEE Transactions on Vehicular Technology}, vol.~71, no.~10, pp. 10\,694--10\,711, Oct. 2022.

\bibitem{8928079}
R.~Li, P.~Gu, and C.~Hua, ``Optimal beam power control for co-existing multibeam {GEO} and {LEO} satellite system,'' in \emph{2019 11th International Conference on Wireless Communications and Signal Processing (WCSP)}, Oct. 2019, pp. 1--6.

\bibitem{9512414}
P.~Gu, R.~Li, C.~Hua, and R.~Tafazolli, ``Dynamic cooperative spectrum sharing in a multi-beam {LEO-GEO} co-existing satellite system,'' \emph{IEEE Transactions on Wireless Communications}, vol.~21, no.~2, pp. 1170--1182, Feb. 2022.

\bibitem{8352660}
C.~Wang, D.~Bian, S.~Shi, J.~Xu, and G.~Zhang, ``A novel cognitive satellite network with {GEO} and {LEO} broadband systems in the downlink case,'' \emph{IEEE Access}, vol.~6, pp. 25\,987--26\,000, 2018.

\bibitem{10273395}
J.~Park, B.~Lee, J.~Choi, H.~Lee, N.~Lee, S.-H. Park, K.-J. Lee, J.~Choi, S.~H. Chae, S.-W. Jeon, K.~S. Kwak, B.~Clerckx, and W.~Shin, ``Rate-splitting multiple access for 6g networks: Ten promising scenarios and applications,'' \emph{IEEE Network}, pp. 1--1, 2023.

\bibitem{9831440}
Y.~Mao, O.~Dizdar, B.~Clerckx, R.~Schober, P.~Popovski, and H.~V. Poor, ``Rate-splitting multiple access: Fundamentals, survey, and future research trends,'' \emph{IEEE Communications Surveys and Tutorials}, vol.~24, no.~4, pp. 2073--2126, Fourthquarter 2022.

\bibitem{10038476}
B.~Clerckx, Y.~Mao, E.~A. Jorswieck, J.~Yuan, D.~J. Love, E.~Erkip, and D.~Niyato, ``A primer on rate-splitting multiple access: Tutorial, myths, and frequently asked questions,'' \emph{IEEE Journal on Selected Areas in Communications}, vol.~41, no.~5, pp. 1265--1308, May 2023.

\bibitem{9451194}
B.~Clerckx, Y.~Mao, R.~Schober, E.~A. Jorswieck, D.~J. Love, J.~Yuan, L.~Hanzo, G.~Y. Li, E.~G. Larsson, and G.~Caire, ``Is {NOMA} efficient in multi-antenna networks? {A} critical look at next generation multiple access techniques,'' \emph{IEEE Open Journal of the Communications Society}, vol.~2, pp. 1310--1343, 2021.

\bibitem{7513415}
H.~Joudeh and B.~Clerckx, ``Robust transmission in downlink multiuser {MISO} systems: {A} rate-splitting approach,'' \emph{IEEE Trans. Signal Processing}, vol.~64, no.~23, pp. 6227--6242, Dec. 2016.

\bibitem{Mao2018}
Y.~Mao, B.~Clerckx, and V.~O. Li, ``Rate-splitting multiple access for downlink communication systems: {B}ridging, generalizing, and outperforming {SDMA and NOMA},'' \emph{Journal on Wireless Communications and Networking}, no. 133 (2018), 2018.

\bibitem{9461768}
Z.~Yang, M.~Chen, W.~Saad, and M.~Shikh-Bahaei, ``Optimization of rate allocation and power control for rate splitting multiple access {(RSMA)},'' \emph{IEEE Transactions on Communications}, vol.~69, no.~9, pp. 5988--6002, Sep. 2021.

\bibitem{8907421}
B.~Clerckx, Y.~Mao, R.~Schober, and H.~V. Poor, ``Rate-splitting unifying {SDMA, OMA, NOMA}, and multicasting in {MISO} broadcast channel: A simple two-user rate analysis,'' \emph{IEEE Wireless Communications Letters}, vol.~9, no.~3, pp. 349--353, March 2020.

\bibitem{7738598}
A.~Zappone, B.~Matthiesen, and E.~A. Jorswieck, ``Energy efficiency in {MIMO} underlay and overlay device-to-device communications and cognitive radio systems,'' \emph{IEEE Transactions on Signal Processing}, vol.~65, no.~4, pp. 1026--1041, Feb. 2017.

\bibitem{8491100}
Y.~Mao, B.~Clerckx, and V.~O. Li, ``Energy efficiency of rate-splitting multiple access, and performance benefits over {SDMA} and {NOMA},'' in \emph{2018 15th International Symposium on Wireless Communication Systems (ISWCS)}, 2018, pp. 1--5.

\bibitem{9123680}
Y.~Mao, B.~Clerckx, J.~Zhang, V.~O.~K. Li, and M.~A. Arafah, ``Max-min fairness of {K}-user cooperative rate-splitting in {MISO} broadcast channel with user relaying,'' \emph{IEEE Transactions on Wireless Communications}, vol.~19, no.~10, pp. 6362--6376, 2020.

\bibitem{9991090}
Y.~Xu, Y.~Mao, O.~Dizdar, and B.~Clerckx, ``Max-min fairness of rate-splitting multiple access with finite blocklength communications,'' \emph{IEEE Transactions on Vehicular Technology}, vol.~72, no.~5, pp. 6816--6821, May 2023.

\bibitem{9970313}
J.~Xu, O.~Dizdar, and B.~Clerckx, ``Rate-splitting multiple access for short-packet uplink communications: A finite blocklength analysis,'' \emph{IEEE Communications Letters}, vol.~27, no.~2, pp. 517--521, Feb 2023.

\bibitem{9562192}
O.~Dizdar, Y.~Mao, Y.~Xu, P.~Zhu, and B.~Clerckx, ``Rate-splitting multiple access for enhanced {URLLC} and {eMBB} in 6{G}: Invited paper,'' in \emph{2021 17th International Symposium on Wireless Communication Systems (ISWCS)}, 2021, pp. 1--6.

\bibitem{liu2022network}
\BIBentryALTinterwordspacing
Y.~Liu, B.~Clerckx, and P.~Popovski, ``Network slicing for {eMBB}, {URLLC}, and {mMTC}: {A}n uplink rate-splitting multiple access approach,'' 2022. [Online]. Available: \url{https://arxiv.org/pdf/2208.10841.pdf}
\BIBentrySTDinterwordspacing

\bibitem{9831048}
Y.~Xu, Y.~Mao, O.~Dizdar, and B.~Clerckx, ``Rate-splitting multiple access with finite blocklength for short-packet and low-latency downlink communications,'' \emph{IEEE Transactions on Vehicular Technology}, vol.~71, no.~11, pp. 12\,333--12\,337, Nov. 2022.

\bibitem{9258414}
W.~Jaafar, S.~Naser, S.~Muhaidat, P.~C. Sofotasios, and H.~Yanikomeroglu, ``On the downlink performance of {RSMA}-based {UAV} communications,'' \emph{IEEE Transactions on Vehicular Technology}, vol.~69, no.~12, pp. 16\,258--16\,263, Dec 2020.

\bibitem{8491094}
M.~Caus, A.~Pastore, M.~Navarro, T.~Ramirez, C.~Mosquera, N.~Noels, N.~Alagha, and A.~I. Perez-Neira, ``Exploratory analysis of superposition coding and rate splitting for multibeam satellite systems,'' in \emph{2018 15th International Symposium on Wireless Communication Systems (ISWCS)}, Aug 2018, pp. 1--5.

\bibitem{9257433}
L.~Yin and B.~Clerckx, ``Rate-splitting multiple access for multigroup multicast and multibeam satellite systems,'' \emph{IEEE Transactions on Communications}, vol.~69, no.~2, pp. 976--990, Feb. 2021.

\bibitem{9684855}
Z.~W. Si, L.~Yin, and B.~Clerckx, ``Rate-splitting multiple access for multigateway multibeam satellite systems with feeder link interference,'' \emph{IEEE Transactions on Communications}, vol.~70, no.~3, pp. 2147--2162, March 2022.

\bibitem{10045781}
H.~Cui, L.~Zhu, Z.~Xiao, B.~Clerckx, and R.~Zhang, ``Energy-efficient {RSMA} for multigroup multicast and multibeam satellite communications,'' \emph{IEEE Wireless Communications Letters}, vol.~12, no.~5, pp. 838--842, May 2023.

\bibitem{9844445}
L.~Yin and B.~Clerckx, ``Rate-splitting multiple access for satellite-terrestrial integrated networks: Benefits of coordination and cooperation,'' \emph{IEEE Transactions on Wireless Communications}, vol.~22, no.~1, pp. 317--332, Jan. 2023.

\bibitem{10266774}
J.~Lee, J.~Lee, L.~Yin, W.~Shin, and B.~Clerckx, ``Coordinated rate-splitting multiple access for integrated satellite-terrestrial networks with super-common message,'' \emph{IEEE Transactions on Vehicular Technology}, pp. 1--6, 2023.

\bibitem{7875081}
Q.~Wang, Z.~Chen, W.~Mei, and J.~Fang, ``Improving physical layer security using{ UAV}-enabled mobile relaying,'' \emph{IEEE Wireless Communications Letters}, vol.~6, no.~3, pp. 310--313, June 2017.

\bibitem{10097680}
W.~U. Khan, Z.~Ali, E.~Lagunas, A.~Mahmood, M.~Asif, A.~Ihsan, S.~Chatzinotas, B.~Ottersten, and O.~A. Dobre, ``Rate splitting multiple access for next generation cognitive radio enabled {LEO} satellite networks,'' \emph{IEEE Transactions on Wireless Communications}, pp. 1--1, 2023.

\bibitem{7765141}
V.~Joroughi, M.~A. Vazquez, and A.~I. Pérez-Neira, ``Generalized multicast multibeam precoding for satellite communications,'' \emph{IEEE Transactions on Wireless Communications}, vol.~16, no.~2, pp. 952--966, Feb. 2017.

\bibitem{6184256}
G.~Zheng, S.~Chatzinotas, and B.~Ottersten, ``Generic optimization of linear precoding in multibeam satellite systems,'' \emph{IEEE Transactions on Wireless Communications}, vol.~11, no.~6, pp. 2308--2320, Jun. 2012.

\bibitem{9628071}
K.-X. Li, L.~You, J.~Wang, X.~Gao, C.~G. Tsinos, S.~Chatzinotas, and B.~Ottersten, ``Downlink transmit design for massive {MIMO LEO} satellite communications,'' \emph{IEEE Transactions on Communications}, vol.~70, no.~2, pp. 1014--1028, Feb. 2022.

\bibitem{9110855}
L.~You, K.-X. Li, J.~Wang, X.~Gao, X.-G. Xia, and B.~Ottersten, ``Massive {MIMO} transmission for {LEO} satellite communications,'' \emph{IEEE Journal on Selected Areas in Communications}, vol.~38, no.~8, pp. 1851--1865, Aug. 2020.

\bibitem{966585}
F.~Fontan, M.~Vazquez-Castro, C.~Cabado, J.~Garcia, and E.~Kubista, ``Statistical modeling of the {LMS} channel,'' \emph{IEEE Transactions on Vehicular Technology}, vol.~50, no.~6, pp. 1549--1567, Nov. 2001.

\bibitem{Bagrov2015}
\BIBentryALTinterwordspacing
A.~V. Bagrov, V.~A. Leonov, A.~S. Mitkin, A.~F. Nasyrov, A.~D. Ponomarenko, K.~M. Pichkhadze, and V.~K. Sysoev, ``Single-satellite global positioning system,'' \emph{Acta Astronautica}, vol. 117, pp. 332--337, 2015. [Online]. Available: \url{https://www.sciencedirect.com/science/article/pii/S0094576515003343}
\BIBentrySTDinterwordspacing

\bibitem{9852292}
L.~You, X.~Qiang, C.~G. Tsinos, F.~Liu, W.~Wang, X.~Gao, and B.~Ottersten, ``Beam squint-aware integrated sensing and communications for hybrid massive {MIMO LEO} satellite systems,'' \emph{IEEE Journal on Selected Areas in Communications}, vol.~40, no.~10, pp. 2994--3009, Oct. 2022.

\bibitem{9149121}
L.~You, K.-X. Li, J.~Wang, X.~Gao, X.-G. Xia, and B.~Otterstenx, ``{LEO} satellite communications with massive {MIMO},'' in \emph{ICC 2020 - 2020 IEEE International Conference on Communications (ICC)}, Jun. 2020, pp. 1--6.

\bibitem{an2021rate}
J.~An, O.~Dizdar, B.~Clerckx, and W.~Shin, ``Rate-splitting multiple access for multi-antenna broadcast channel with imperfect {CSIT} and {CSIR},'' in \emph{IEEE International Symposium on Personal, Indoor and Mobile Radio Communications (PIMRC)}, 2020.

\bibitem{9165811}
J.~Chu, X.~Chen, C.~Zhong, and Z.~Zhang, ``Robust design for {NOMA}-based multibeam {LEO} satellite internet of things,'' \emph{IEEE Internet of Things Journal}, vol.~8, no.~3, pp. 1959--1970, Feb. 2021.

\bibitem{grant2008cvx}
M.~Grant, S.~Boyd, and Y.~Ye, ``{CVX}: Matlab software for disciplined convex programming,'' 2008.

\bibitem{7555358}
H.~Joudeh and B.~Clerckx, ``Sum-rate maximization for linearly precoded downlink multiuser {MISO} systems with partial {CSIT}: {A} rate-splitting approach,'' \emph{IEEE Transactions on Communications}, vol.~64, no.~11, pp. 4847--4861, Nov. 2016.

\bibitem{8642812}
Z.~Lin, M.~Lin, J.-B. Wang, T.~de~Cola, and J.~Wang, ``Joint beamforming and power allocation for satellite-terrestrial integrated networks with non-orthogonal multiple access,'' \emph{IEEE Journal of Selected Topics in Signal Processing}, vol.~13, no.~3, pp. 657--670, June 2019.

\bibitem{9180053}
G.~Zhou, C.~Pan, H.~Ren, K.~Wang, and A.~Nallanathan, ``A framework of robust transmission design for {IRS}-aided {MISO} communications with imperfect cascaded channels,'' \emph{IEEE Transactions on Signal Processing}, vol.~68, pp. 5092--5106, 2020.

\bibitem{ben2001lectures}
A.~Ben-Tal and A.~Nemirovski, \emph{Lectures on modern convex optimization: analysis, algorithms, and engineering applications}.\hskip 1em plus 0.5em minus 0.4em\relax SIAM, 2001.

\end{thebibliography}

\end{document}